\documentclass[12pt]{article}
\usepackage{cite}
\usepackage{graphicx}
\usepackage{amssymb,amsfonts}
\usepackage{hangcaption}

\renewcommand{\theequation}
{\arabic{section}.\arabic{equation}}

\makeatletter
\def\eqnarray{ \stepcounter{equation} \let\@currentlabel=\theequation
 \global\@eqnswtrue
 \global\@eqcnt\z@
 \tabskip\@centering
 \let\\=\@eqncr
 $$\halign to \displaywidth\bgroup\@eqnsel\hskip\@centering
 $\displaystyle\tabskip\z@{##}$&\global\@eqcnt\@ne
 \hfil$\displaystyle{{}##{}}$\hfil
 &\global\@eqcnt\tw@$\displaystyle\tabskip\z@{##}$\hfil
 \tabskip\@centering&\llap{##}\tabskip\z@\cr}
\makeatother

\makeatletter
\def\@arrayacol{\edef\@preamble{\@preamble \hskip .5\arraycolsep}}
\def\array{\let\@acol\@arrayacol \let\@classz\@arrayclassz
\let\@classiv\@arrayclassiv \let\\\@arraycr\def\@halignto{}\@tabarray}
\makeatother

\renewcommand{\arraystretch}{1.6}

\makeatletter
\newcounter{subeqncnt}
\def\thesubeqncnt{\alph{subeqncnt}}
\def\subequations{\begingroup%
   \stepcounter{equation}\edef\@tempa{\theequation}%
   \let\c@equation\c@subeqncnt\c@subeqncnt\z@
   \edef\theequation{\@tempa\noexpand\thesubeqncnt}}

\makeatother

\newcommand{\captionfonts}{\small} 
\makeatletter 
\long\def\@makecaption#1#2{%
\vskip\abovecaptionskip
\sbox\@tempboxa{{\captionfonts #1: #2}}%
\ifdim \wd\@tempboxa >\hsize
{\captionfonts #1: #2\par}
\else
\hbox to\hsize{\hfil\box\@tempboxa\hfil}%
\fi
\vskip\belowcaptionskip}
\makeatother 

\newcommand{\del}{\partial}

\newcommand{\tr}{{\rm Tr}\,}

\newcommand{\dd}{{\rm d}}

\newcommand{\const}{{\rm const.}}

\begin{document}

\setlength{\baselineskip}{7mm}
\begin{titlepage}
\begin{flushright}
{\tt APCTP-Pre2008-002} \\
{\tt arXiv:0806.4460[hep-th]} \\
June, 2008
\end{flushright}

\vspace{1cm}

\begin{center}

{\Large Density Dependence of Transport Coefficients \\
from Holographic Hydrodynamics}

\vspace{1cm}

{\sc{Xian-Hui Ge}}$^*$,
{\sc{Yoshinori Matsuo}}$^*$,
{\sc{Fu-Wen Shu}}$^*$,  \\
{\sc{Sang-Jin Sin}}$^\dagger{}^*$
and
{\sc{Takuya Tsukioka}}$^*$

\vspace*{3mm}

$*$
{\it{Asia Pacific Center for Theoretical Physics},} \\
{\it{Pohang, Gyeongbuk 790-784, Korea}} \\
{\sf{gexh, ymatsuo, fwshu, tsukioka@apctp.org}}
\\
$\dagger$
{\it{Department of Physics,}}
{\it{Hanyang University,}} 
{\it{Seoul 133-791, Korea}} \\
{\sf{sjsin@hanyang.ac.kr}}
\end{center}

\vspace{1cm}

\begin{abstract}
We study the transport coefficients of Quark-Gluon-Plasma 
in finite temperature and finite baryon density. 
We use AdS/QCD of charged AdS black hole background 
with bulk-filling branes identifying 
the $U(1)$ charge as the baryon number.  
We calculate the diffusion constant, the shear viscosity 
and the thermal conductivity to plot their density and 
temperature dependences.
Hydrodynamic relations between those are shown to hold exactly.   
The diffusion constant and the shear viscosity are decreasing 
as a function of density for fixed total energy.  
For fixed temperature, the fluid becomes less diffusible 
and more viscous for larger baryon density. 
\end{abstract}

\end{titlepage}

\section{Introduction}
\setcounter{equation}{0}
\setcounter{footnote}{0}

After the discovery of consistency on the ratio of the viscosity 
to the entropy density $(\eta/s)$~\cite{pss0} in 
AdS/CFT correspondence~\cite{ads/cft, gkp, w}  
and RHIC (Relativistic Heavy Ion Collider) experiment, 
much attention has been drawn to the calculational 
scheme provided by string theory.
Even some attempt has been made to map the entire process 
of RHIC experiment in terms of the gravity dual~\cite{ssz}.  
The way to include a chemical potential in the theory was 
figured out in the context of 
D4D8$\overline{\mbox{D8}}$ setup~\cite{ksz,ht}. 
Phases of these theories were discussed in D3/D7 setup and 
new phases were reported where instability due to the strong 
attraction is a feature~\cite{nssy1,kmmmt,nssy2}. 

Although QCD and ${\cal N}=4$ SYM are different, 
it is expected that some of the properties are shared 
by the two theories. 
It is an interesting question to ask 
how much one can learn by studying the various versions 
of AdS/CFT correspondence.  
The relevance is based on the universality of low energy physics.
In this respect, the hydrodynamic limit is interesting 
since such limit can be shared by many theories in spite of 
the differences in UV limit. 

The calculation scheme of transport coefficients 
is to use Kubo formula, which gives a relation to the low energy 
limit of Wightman Green functions.
In AdS/CFT correspondence, one calculate the retarded 
Green function which is related  to 
the Wightman function by fluctuation-dissipation theorem.
Such scheme has been developed 
in a series of papers~\cite{ss,pss,pss2,hs,ks}.

For charged case, the calculations are more involved 
and corresponding works have been done only partially 
by various groups~\cite{mas,ss2,mno, bbn}.
In~\cite{mas,ss2}, full thermodynamics for STU black 
hole~\cite{cvetic,cvetic2} and the hydrodynamic calculations 
were performed for the $(1,0,0)$ charge. 
In~\cite{mno}, charged  AdS black hole background was 
considered and it was shown  numerically that the ratio $(\eta/s)$ 
was $1/(4\pi)$ with very good accuracy. 
Later, it was also proven that the ratio was universal in more general 
setup~\cite{bbn}. 

In this paper, we perform hydrodynamic calculation directly for  
Reissner-Nordstr{\"o}m-Anti-deSitter (RN-AdS) black hole, 
which corresponds to the $(1,1,1)$ charged STU black hole. 
Master equations for the decoupled modes are worked out explicitly. 
Vector modes of Maxwell field (as well as the vector modes of metric) 
have diffusion pole, contrary to the chargeless case.   
As a consequence,  the diffusion constant is calculated and 
the ratio $(\eta/s)=1/(4\pi)$ is shown to be exact in this case also. 
We observe that the diffusion constant and the shear viscosity 
decrease as we increase the charge with fixed total system energy 
(or equivalently the black hole mass), 
while the shear viscosity increases for fixed temperature. 

The charge in RN-AdS black hole is usually regarded 
as $R$-charge of SUSY~\cite{myers}. 
We here consider an another interpretation in the following way:   
One can introduce quarks and mesons by considering 
the bulk-filling branes in AdS$_5$ space. 
The overall $U(1)$ of the flavor branes is identified 
as the baryon charge. 
The $U(1)$ charge in this model~\cite{s} minimally couples 
to the bulk gravity since the bulk and the world volume of brane  
are identified.
RN-AdS metric can be considered 
as the consequence of the back reaction of the AdS 
black hole to the charge.  
Therefore the $U(1)$ charge in RN-AdS can be identical to the baryon 
charge.
As a result, we can calculate 
the transport coefficients in the presence of the baryon density. 

One can give an explanation of hydrodynamic mode in meson physics. 
In our interpretation, the Maxwell fields are the fluctuations of 
bulk-filling branes, therefore they should be interpreted as
master fields of the mesons. 
Then hydrodynamic modes are lowest lying massless meson spectrum. 
In terms of brane embedding picture, 
this massless-ness is due to the touching of the brane 
on the black hole horizon. 
Near the horizon, the tension of the brane is zero due to 
the metric factor and it can lead to the massless fluctuation. 
Then the massless spectrum can not go far from the horizon 
in radial direction.  
In this picture, hydrodynamic nature is closely related 
to the near horizon behavior of the branes. 

This paper is organized as follows: 
In section~\ref{sec:Setup}, we introduce RN-AdS black hole and 
review correlation function calculation at finite temperature 
in AdS/CFT correspondence. 
In section~\ref{sec:Perturb}, 
a formulation on the metric and the gauge perturbations  
in RN-AdS background is given.  
We then calculate the correlators in hydrodynamic regime and obtain 
the diffusion pole in section~\ref{sec:Diffusion}. 
In section~\ref{sec:Viscosity}, 
the shear viscosity is calculated via Kubo formula. 
We also show that the  result is consistent with the hydrodynamic 
relation of diffusion constant and the viscosity. 
The thermal conductivity is also calculated in this section.   
Conclusions and discussions are given in section~\ref{sec:Conclusion}. 
Three appendices are given to provide the details of the calculations.

\section{Basic Setup}\label{sec:Setup}
 
\subsection{Minkowskian correlators in AdS/CFT correspondence}
\setcounter{equation}{0}
\setcounter{footnote}{0}

Before introducing RN-AdS black hole, 
we briefly summarize Minkowskian correlators 
in AdS/CFT correspondence. 
We follow the prescription proposed in~\cite{ss}.
Let us consider fluctuations of fields which satisfy 
equations of motion at the linearized order. 
We work on the five-dimensional background,  
\begin{equation}
 \dd s^2 = g_{\mu\nu}\dd x^\mu\dd x^\nu + g_{uu}(\dd u)^2,  
\end{equation}
where $x^\mu$ and $u$ are the four-dimensional and the radial coordinates, 
respectively. 
We refer the boundary as $u=0$ and the horizon as $u=1$.  
A solution of the equation of motion may be given,  
\begin{equation}
\phi(u,x) = 
\!\int\!\frac{\dd^4 k}{(2\pi)^4}\ \mbox{e}^{ikx}f_k(u)\phi_0(k), 
\end{equation}
where $f_k(u)$ is normalized such that $f_k(0)=1$ at the boundary.  
An on-shell action might be reduced to surface terms 
by using the equation of motion, 
\begin{equation}
S[\phi_0]
=\!\int\!\frac{\dd^4 k}{(2\pi)^4}
\phi_0(-k){\cal G}(k, u)\phi_0(k)
\bigg|_{u=0}^{u=1} . 
\label{on_shell_action}
\end{equation}
Here, the function $\mathcal G(k,u)$ 
can be written in terms of $f_{\pm k}(u)$ and $\partial_u f_{\pm k}(u)$, 
for example, for a scalar field, 
$$
 \mathcal G(k,u) 
= K \sqrt{-g} g^{uu} f_{-k}(u) \partial_u f_k(u),  
$$
with some constant $K$. 
The direct generalization of AdS/CFT correspondence, 
or Gubser-Klebanov-Polyakov/Witten relation~\cite{gkp,w}, 
to Minkowski spacetime gives the relation,  
$$
 \left\langle \mbox{e}^{i\!\int\!\phi_0\mathcal O}\right\rangle 
= \mbox{e}^{iS[\phi_0]},  
$$
where the operator ${\cal O}$ is defined in the boundary field theory. 
From this relation, one may obtain a Green function 
by taking second derivative of the action with respect to the boundary 
value of the field, 
$$
 G(k) = -\mathcal G(k,u)\biggr|^{u=1}_{u=0}
  -\mathcal G(-k,u)\biggr|^{u=1}_{u=0}. 
$$
However, this quantity is real 
and cannot be a retarded Green function. 
This can be seen as follows. 
The imaginary part of $\mathcal G(k,u)$ 
is proportional to a conserved flux. 
Then, its contributions at the boundary $u=0$ 
and at the horizon $u=1$ cancel completely. 
Even if one neglects the contribution from the horizon, 
$G(k)$ is still real. 
The reality condition of the equation of motion 
implies $\mathcal G(-k,u) = \mathcal G^*(k,u)$,  
and the imaginary part of $G(k)$ vanishes again. 
Therefore we should impose the ``retarded'' condition to 
the Green function. 

Son and Starinets proposed that 
the retarded (advanced) Green function is given by 
\begin{equation}
G^{\rm R}(k)
=
2{\cal G}(k, u)
\bigg|_{u=0}, 
\label{green_function}
\end{equation}
with incoming (outgoing) boundary condition at the horizon. 
Generally, the contribution at the horizon 
is oscillating and averaged out to zero. 
In order to avoid this, we have to consider 
incoming or outgoing boundary condition. 
Taking away the contribution at the horizon, 
we obtain $\mathcal G(k,u)$ with a non-zero imaginary part. 
Physics at the horizon affects the Green function 
only through the boundary condition. 
In general, there are several fields in the model. 
We write the Green function as $G_{ij}(k)$, 
where indices $i$ and $j$ distinguish these fields. 
The surface terms are always associated with 
equations of motion. 
We choose the former index to indicate the field 
whose equation of motion is associated with the Green function. 

In this paper, we work in RN-AdS background and 
consider its perturbations so that 
essential ingredients are perturbed metric field and $U(1)$ 
gauge field.  
Here we define the precise form of the retarded Green function 
which we discuss later:   
%
\renewcommand{\arraystretch}{2.0}
%
\begin{eqnarray}
\begin{array}{rcl}
G_{\mu\nu \ \rho\sigma}(k)
&=&
\displaystyle
-i\!\int\!\dd^4x
\ \mbox{e}^{-ikx}\theta(t)
\langle{[}T_{\mu\nu}(x), \ T_{\rho\sigma}(0){]}\rangle, 
\\
G_{\mu\nu \ \rho}(k)
&=&
\displaystyle
-i\!\int\!\dd^4x
\ \mbox{e}^{-ikx}\theta(t)
\langle{[}T_{\mu\nu}(x), \ J_\rho(0){]}\rangle, 
\\
G_{\mu \ \nu}(k)
&=&
\displaystyle
-i\!\int\!\dd^4x
\ \mbox{e}^{-ikx}\theta(t)
\langle{[}J_\mu(x), \ J_\nu(0){]}\rangle, 
\end{array}
\label{diff_green_function}
\end{eqnarray}
%
\renewcommand{\arraystretch}{1.6}
%

\vspace*{-3mm}
\noindent
where the operators $T_{\mu\nu}(x)$ and $J_\mu(x)$ are energy-momentum 
tensor and $U(1)$ current which couple to the metric and the gauge fields,  
respectively.

\subsection{Reissner-Nordstr{\"o}m-AdS background}

In this paper, we consider $N_c$ D3-branes 
and $N_f$ D7-branes, and treat the D3-branes as a gravitational background. 
The D7-branes are wrapping on $S^3$ of $S^5$, 
and we neglect this $S^3$ dependence. 
We do not consider the perpendicular 
fluctuations of D7-branes, 
and the effective action then becomes that for  
five-dimensional gauge theory. 
If the D7-branes touch the D3-branes, 
the D7-branes fill the AdS${}_5$ completely.
The induced metric on the D7-brane 
is identical to the AdS bulk metric\cite{s}.
This model corresponds to ${\cal N}=4$ SYM with massless quarks. 
If we introduce the baryon charge at the boundary theory, 
its chemical potential is identified as the tail of the $U(1)$ 
gauge potential on the flavor brane\cite{ksz,ht}. 
(See also \cite{nssy1,kmmmt,nssy2,bergman,ubc} for later development.) 
We consider the phenomenological model taking only 
AdS${}_5$ part and neglecting $S^5$ part.  
Then there is no way to distinguish the bulk gauge field and 
the brane field. 
The baryon charge and the $R$-charge have the same description in terms 
of the $U(1)$ gauge field living in the AdS${}_5$ space.   
A charged black hole (RN-AdS black hole) is then induced by its back 
reaction. 
This corresponds to the ${\cal N}=4$ SYM 
in finite temperature with finite baryon density. 
It is the case that we consider in this paper.

The action for the gauge field dual to the baryon current is given 
by the $U(1)$ part of the Dirac-Born-Infeld action\footnote{
The indices $m$ and $n$ run through five-dimensional spacetime
while $\mu$ and $\nu$ would be reserved for four-dimensional Minkowski
spacetime. Their spatial coordinates are labeled by $i$ and $j$.
}
\begin{equation}
 S_{\mathrm{D7}} 
= 
-\frac{1}{4e^2}
\!\int\!\dd^5x\sqrt{-g}\ 
\tr\bigl({\cal F}_{mn}{\cal F}^{mn}\bigl) ,  
\end{equation}
where the gauge coupling constant $e$ is given by 
\begin{equation}
 \frac{l}{e^2} = \frac{N_cN_f}{(2\pi)^2} , 
\end{equation}
with $l$ the radius of the AdS space.  
Notice that the  gauge field is that of the diagonal $U(1)$ 
of $U(N_f)$ flavor brane dynamics, 
which is dual to the baryon current at the boundary. 
Together with the gravitation part, 
we arrive at the following action which is our starting point:  
\begin{equation}
S[g_{mn}, {\cal A}_m]
=\frac{1}{2\kappa^2}\!\int\!\dd^5x\sqrt{-g}
\Big(
R-2\Lambda
\Big)
-
\frac{1}{4e^2}
\!\int\!\dd^5x\sqrt{-g}
{\cal F}_{mn}{\cal F}^{mn},
\label{action_bh}
\end{equation}
where we denote the gravitation constant and the cosmological constant 
as $\kappa^2=8\pi G_5$ and $\Lambda$, respectively.  
The $U(1)$ gauge field strength is given by 
${\cal F}_{mn}(x)=\del_m{\cal A}_n(x)-\del_n{\cal A}_m(x)$. 
The gravitation constant is related to the gauge theory quantities by 
\begin{equation}
 \frac{l^3}{\kappa^2} = \frac{N_c^2}{4\pi^2} . 
\end{equation}
Suppose we have baryon charge $Q$. 
This should be identified to the source of $U(1)$ charge on the brane 
hence on the bulk.
Then we can relate it to the 
parameter in RN black hole solution by considering the
full solution to the equation of motion, 
\begin{equation}
R_{mn}-\frac{1}{2}g_{mn}R+g_{mn}\Lambda
=
\kappa^2T_{mn},
\label{eq_motion_bh}
\end{equation}
where energy-momentum tensor $T_{mn}(x)$ is given by
\begin{equation}
T_{mn}
=\frac{1}{e^2}
\bigg(
{\cal F}_{mk}{\cal F}_{nl}g^{kl}
-\frac{1}{4}g_{mn}{\cal F}_{kl}{\cal F}^{kl}
\bigg).
\end{equation}
An equation of motion for the gauge field 
${\cal A}_m(x)$ gives Maxwell equation, 
\begin{equation}
 \nabla_m{\cal F}^{mn}
=\frac{1}{\sqrt{-g}}\del_m
\Big(\sqrt{-g}g^{mk}g^{nl}
\big(\del_k{\cal A}_l-\del_l{\cal A}_k\big)\Big)=0.
\label{maxwell_eq_0}
\end{equation} 
Here we assumed that there is no electromagnetic source outside 
the black hole. 
One can confirm that the following metric and gauge potential
satisfy the equations of motion (\ref{eq_motion_bh}) and 
(\ref{maxwell_eq_0}),
\begin{subequations}
\begin{eqnarray}
\dd s^2
&=&
\frac{r^2}{l^2}
\bigg(
-f(r)(\dd t)^2+\sum_{i=1}^3(\dd x^i)^2
\bigg)
+\frac{l^2}{r^2f(r)}(\dd r)^2,
\label{rnads}
\\
{\cal A}_t
&=&
-\frac{Q}{r^2}+\mu,
\label{rnads_1}
\end{eqnarray}
\end{subequations}

\vspace*{-7mm}
\noindent
with
$$
f(r)
=
1-\frac{ml^2}{r^4}+\frac{q^2l^2}{r^6},
\qquad
\Lambda
=
-\frac{6}{l^2}, \qquad
$$
if and only if $q$ is related to the $Q$ by
\begin{equation}
 e^2=\frac{2Q^2}{3q^2}\kappa^2. 
\end{equation}
It should be noted that 
a ratio of the gauge coupling constant $e^2$ to 
the gravitation constant $\kappa^2$ is 
\begin{equation}
\frac{e^2}{\kappa^2}
=\frac{N_c}{N_f}l^{-2}.
\end{equation}
Since the gauge potential ${\cal A}_t(x)$ must vanish at the horizon, 
the charge $Q$ and the chemical potential $\mu$ are related.  
The parameters $m$ and $q$ are 
the mass and charge of AdS space, respectively.
This is nothing but Reissner-Nordstr{\"o}m-Anti-deSitter 
(RN-AdS) background
in which we are interested throughout this paper.

The horizons of RN-AdS black hole are located
at the zero for $f(r)$\footnote{
In order to define the horizon, the charge $q$ must satisfy
a relation $q^4\le 4m^3l^2/27$.
},
\begin{equation}
f(r)
=
1-\frac{ml^2}{r^4}+\frac{q^2l^2}{r^6}
=
\frac{1}{r^6}
\Big(r^2-r_+^2\Big)
\Big(r^2-r_-^2\Big)
\Big(r^2-r_0^2\Big), 
\label{metric} 
\end{equation}
where their explicit forms of the horizon radiuses are given by
\begin{subequations}
\begin{eqnarray}
r_+^2
&=&
\left(
\frac{m}{3q^2}
\Bigg(
1+2\cos\bigg(\frac{\theta}{3}+\frac{4}{3}\pi\bigg)
\Bigg)
\right)^{-1},
\label{r+}
\\
r^2_-
&=&
\left(
\frac{m}{3q^2}
\Bigg(
1+2\cos\bigg(\frac{\theta}{3}\bigg)
\Bigg)
\right)^{-1},
\label{r-}
\\
r_0^2
&=&
\left(
\frac{m}{3q^2}
\Bigg(
1+2\cos
\bigg(
\frac{\theta}{3}
+\frac{2}{3}\pi
\bigg)
\Bigg)
\right)^{-1},
\label{r0}
\end{eqnarray}
\end{subequations}

\vspace*{-7mm}
\noindent
with
$$
\theta
=
\arctan
\Bigg(
\frac{3\sqrt{3}q^2\sqrt{\displaystyle 4m^3l^2-27q^4}}{2m^3l^2-27q^4}
\Bigg),
$$
and satisfy a relation $r^2_++r^2_-=-r^2_0$.
The positions expressed by $r_+$ and $r_-$ correspond to the outer
and the inner horizon, respectively. 
It will be useful to notice that the charge 
$q$ can be expressed in terms of $\theta$ and $m$ by
$$
q^4=\frac{4m^3l^2}{27}\sin^2\left(\frac{\theta}{2}\right).
$$
The outer horizon takes a value in
$$
\sqrt{\frac{m}{3}}l
\le r_+^2 \le \sqrt{m}l,
$$
where the upper bound and the lower bound correspond to the case for 
$q=0$ and the extremal case, respectively.  

We shall give various thermodynamic quantities of RN-AdS black
hole~\cite{myers, s}.
The temperature is defined from the conical singularity free
condition around the horizon $r_+$,
\begin{equation}
T
=
\frac{r_+^2f'(r_+)}{4\pi l^2}
=
\frac{r_+}{\pi l^2}
\bigg(
1-\frac{1}{2}\frac{q^2l^2}{ r_+^6}
\bigg)
\equiv
\frac{1}{2\pi b}
\Big(
1-\frac{a}{2}  
\Big),
\quad (>0),
\label{temp}
\end{equation}
where we defined the parameters $a$ and $b$ by
\begin{equation}
a\equiv\frac{q^2l^2}{r_+^6}, \qquad
b\equiv\frac{l^2}{2r_+}. 
\end{equation}
In the limit $q\rightarrow 0$,  these parameters go to 
$$
a\rightarrow 0,  \qquad 
b\rightarrow  \frac{l^{3/2}}{2m^{1/4}},
$$
and the temperature becomes 
\begin{equation}
T\rightarrow T_0
= \frac{m^{1/4}}{\pi l^{3/2}}.
\label{temp_0} 
\end{equation}
The entropy density $s$, the energy density $\epsilon$, the
pressure $p$, the chemical potential $\mu$ and the density of physical 
charge $\rho$ can be also computed as
\begin{eqnarray}
s
&=&
\frac{2\pi r_+^3}{\kappa^2l^3}
=\frac{\pi l^3}{4\kappa^2b^3},
\label{entropy}
\\
\epsilon
&=&
\frac{3m}{2\kappa^2l^3}
=\frac{3l^3}{32\kappa^2b^4}\Big(1+a\Big), 
\label{energy}
\\
p
&=&
\frac{\epsilon}{3}, 
\label{pressure} 
\\
\mu
&=&
\frac{Q}{r_+^2}, 
\label{chemical_potential} 
\\
\rho
&=&
\frac{2Q}{e^2l^3}.
\label{charge_density}
\end{eqnarray}
%

\section{Perturbations in RN-AdS Background}\label{sec:Perturb}
\setcounter{equation}{0}
\setcounter{footnote}{0}

In RN-AdS background, we study small 
perturbations of the metric $g_{mn}(x)$ and the gauge 
field ${\cal A}_m(x)$,  
\begin{equation}
\begin{array}{rcl}
g_{mn}
&\equiv&
g^{(0)}_{mn}+h_{mn}, 
\\
{\cal A}_m
&\equiv&
A_m^{(0)}+A_m,  
\end{array}
\end{equation}
where the background metric $g^{(0)}_{mn}(x)$ and 
the background gauge field $A^{(0)}_m(x)$ 
are given in (\ref{rnads}) and (\ref{rnads_1}), 
respectively. 
In the metric perturbation, one can define a inverse metric as
$$
g^{mn}=g^{(0)mn}-h^{mn} + {\cal O}(h^2),
$$
and raise and lower indices by using the background metric
$g_{mn}^{(0)}(x)$ and $g^{(0)mn}(x)$.

Now we shall consider a linearized theory 
of the symmetric tensor field $h_{mn}(x)$ 
and the vector field $A_m(x)$  
propagating in RN-AdS background.
In the first order of $h_{mn}(x)$ and $A_m(x)$, 
the Einstein equation (\ref{eq_motion_bh}) can be written as
\begin{equation}
R^{(1)}_{mn}
-\frac{1}{2}g^{(0)}_{mn}R^{(1)}
-\frac{1}{2}h_{mn}R^{(0)}
+h_{mn}\Lambda
=\kappa^2
T^{(1)}_{mn}.
\label{einstein_eq_bh_01}
\end{equation}
In the expression above,
the scalar curvature $R^{(0)}(x)$ 
is constructed by using the background
metric $g_{mn}^{(0)}(x)$
and the following tensors are newly defined: 
\begin{eqnarray*}
R^{(1)}_{mn}
&=&
\frac{1}{2}
\Big(
\nabla_k\nabla_m h_n{}^k
+\nabla_k\nabla_n h_m{}^k
-\nabla_k\nabla^kh_{mn}
-\nabla_m\nabla_nh
\Big),
\\
R^{(1)}
&=&
g^{(0)kl}R_{kl}^{(1)}-h^{kl}R_{kl}^{(0)}
\nonumber
\\
&=&
\nabla_{k}\nabla_{l}h^{kl}
-\nabla_k\nabla^kh
-h^{kl}R_{kl}^{(0)},
\\
T^{(1)}_{mn}
&=&
\frac{1}{e^2}
\Big(
-F_{mk}^{(0)}F_{nl}^{(0)}h^{kl}
+\frac{1}{2}g^{(0)}_{mn}F_{kp}^{(0)}F^{(0)}{}_l{}^ph^{kl}
-\frac{1}{4}h_{mn}F_{kl}^{(0)}F^{(0)kl} 
\nonumber 
\\
&&
\hspace*{8mm} 
+F^{(0)}_{mk}F_n{}^k
+F^{(0)}_{nk}F_m{}^k 
-\frac{1}{2}g_{mn}^{(0)}F_{kl}^{(0)}F^{kl}
\Big),
\end{eqnarray*}
where the Ricci tensor $R^{(0)}_{mn}(x)$, the covariant derivative 
and the field strength $F^{(0)}_{mn}(x)$  
are defined through the background metric $g^{(0)}_{mn}(x)$ and 
the gauge field $A^{(0)}_m(x)$. 
We denote a trace part of the metric and a field strength 
for the perturbative 
parts as $h(x)\equiv h_{mn}g^{(0)mn}(x)$ and 
$F_{mn}(x)\equiv \del_mA_n(x)-\del_nA_m(x)$, respectively.   
On the other hand, 
the Maxwell equation (\ref{maxwell_eq_0}) becomes 
\begin{eqnarray}
0
&=&
\nabla_m
\bigg(
F^{mn}
-F^{(0)}{}^m{}_kh^{nk}
+F^{(0)}{}^n{}_kh^{mk}
+\frac{1}{2}F^{(0)mn}h
\bigg)
\nonumber 
\\
&=&
\frac{1}{\sqrt{-g^{(0)}}}
\del_m\bigg\{\sqrt{-g^{(0)}}
\bigg(
g^{(0)mk}g^{(0)nl}(\del_kA_l-\del_lA_k)
\nonumber 
\\
&&
\hspace*{40mm}
-F^{(0)}{}^m{}_kh^{nk}
+F^{(0)}{}^n{}_kh^{mk}
+\frac{1}{2}F^{(0)mn}h
\bigg)
\bigg\}.
\label{maxwell_eq_01}
\end{eqnarray}
The above equations of motion (\ref{einstein_eq_bh_01}) and 
(\ref{maxwell_eq_01}) can be derived from the following
action:
\begin{eqnarray}
S[h_{mn}, A_m]
=
&&
-\frac{1}{4\kappa^2}
\!\int\!\dd^5x\sqrt{-g^{(0)}}
\Bigg\{
\nabla_mh^{mn}\nabla_nh
-\nabla_mh^{nk}\nabla_nh^m{}_k
\nonumber
\\
&&
\hspace*{38mm}
+\frac{1}{2}\nabla_mh^{kl}\nabla^mh_{kl}
-\frac{1}{2}\nabla_mh\nabla^mh
\nonumber
\\
&&
\hspace*{38mm}
+
\bigg(
\frac{1}{2}R^{(0)}-\Lambda-\frac{\kappa^2}{4e^2}F^{(0)}_{kl}F^{(0)kl}
\bigg)
\Big(
\frac{1}{2}h^2-h_{mn}h^{mn}
\Big)
\nonumber
\\
&&
\hspace*{38mm}
+\frac{\kappa^2}{e^2}F^{(0)}_{mn}F^{(0)}_{kl}h^{mk}h^{nl}
\Bigg\}
\nonumber 
\\
&&
-\frac{1}{4e^2}
\!\int\!\dd^5x\sqrt{-g^{(0)}}
\Bigg\{
F_{mn}F^{mn}
\nonumber 
\\
&&
\hspace*{36mm}
-2
\bigg(
F^{(0)}_{mk}F_n{}^kh^{mn}
+F^{(0)}_{nk}F_m{}^kh^{mn}
-\frac{1}{2}F^{(0)}_{mn}F^{mn}h
\bigg)
\Bigg\}.
\nonumber 
\\
\end{eqnarray}
By using the equations of motion, 
an on-shell action is reduced to surface term
\begin{eqnarray}
S[h_{mn}^{\rm cl}, A_m^{\rm cl}]
&=&
-\frac{1}{8\kappa^2}\!\int\!\dd^5 x\del_m
\Bigg\{
\sqrt{-g^{(0)}}
\bigg(
h^{mn}\nabla_nh
+h\nabla_nh^{mn}
-2h^{nk}\nabla_nh^m{}_k
\nonumber
\\
&&
\hspace*{46mm}
+h^{kl}\nabla^mh_{kl}
-h\nabla^mh
\bigg)
\Bigg\}
\nonumber 
\\
&&
-\frac{1}{2e^2}
\!\int\!\dd^5x\del_m
\Bigg\{
\sqrt{-g^{(0)}}
A_n\bigg(
F^{mn}
\nonumber 
\\
&&
\hspace*{50mm}
-
F^{(0)m}{}_kh^{nk}
+F^{(0)}{}^n{}_kh^{mk}
+\frac{1}{2}F^{(0)mn}h
\bigg)
\Bigg\}.
\nonumber 
\\
\label{on-shell_action_h_a}
\end{eqnarray}

We shall work in the $h_{rm}(x)=0$ and $A_r(x)=0$ gauges and 
use the Fourier
decomposition
\begin{eqnarray*}
h_{\mu\nu}(t, z, r)
&=&
\!\int\!\frac{\dd^4k}{(2\pi)^4}
\ \mbox{e}^{-i\omega t+ikz}h_{\mu\nu}(k, r), 
\\ 
A_\mu(t, z, r)
&=&
\!\int\!\frac{\dd^4k}{(2\pi)^4}
\ \mbox{e}^{-i\omega t+ikz}
A_\mu(k, r), 
\end{eqnarray*}
where we choose the momenta which are along the $z$-direction.  
In this case,
one can categorize the metric perturbations to the following three
types by using the spin under the $O(2)$ rotation 
in $(x, y)$-plane~\cite{pss}:
\begin{itemize}
\item vector type: \
$h_{xt}\ne0$, \ $h_{xz}\ne0$, \ ${\mbox{(others)}}=0$
\\
\hspace*{27mm}
$\Big($equivalently, $h_{yt}\ne0$, \ $h_{yz}\ne0$, \
${\mbox{(others)}}=0$$\Big)$
\item tensor type: \
$h_{xy}\ne0$, \ $h_{xx}=-h_{yy}\ne0$, \ ${\mbox{(others)}}=0$
\item scalar type: \
$h_{tz}\ne0$, $h_{tt}\ne0$, $h_{xx}=h_{yy}\ne0$, and $h_{zz}\ne0$,
\ $\mbox{(others)}=0$
\end{itemize}
We consider the first two types in this paper.
The scalar type perturbation would be studied elsewhere.

\subsection{Vector type perturbation}

In this subsection, we study the vector type 
perturbation in RN-AdS background.
From explicit calculation, one can show that only $x$-component of the 
gauge field $A_x(x)$ could participate in the linealized 
perturbative equations of motion. 
Thus independent variables are 
$$
h_{xt}(x)\ne 0, \quad  
h_{xz}(x)\ne 0, \quad
A_{x}(x)\ne 0, \quad
 (\mbox{others})=0.
$$
We start by introducing new field valiables,   
$h^x_t(r)=g^{(0)xx}h_{xt}(r)=(l^2/r^2)h_{xt}(r)$ and 
$h^x_z(r)=g^{(0)xx}h_{xz}(r)=(l^2/r^2)h_{xz}(r)$.  
Nontrivial equations in the Einstein equation (\ref{einstein_eq_bh_01}) 
appear from $(t, x)$, $(r, x)$ and $(x, z)$ components, respectively: 
\begin{subequations}
\begin{eqnarray}
0
&=&
{h^x_t}''
+\frac{5}{r}{h^x_t}'
-\frac{l^4}{r^4f}\Big(\omega kh^x_z+k^2h^x_t\Big)
+\frac{6q^2l^2}{Qr^5} A_x', 
\label{eq_motion_v_01} 
\\
0
&=&
kf{h^x_z}'
+\omega{h^x_t}'
+\frac{6q^2l^2\omega}{Qr^5}A_x, 
\label{eq_motion_v_02}
\\
0
&=&
{h^x_z}''
+\frac{(r^5f)'}{r^5f}{h^x_z}'
+\frac{l^4}{r^4f^2}\Big(\omega kh^x_t+\omega^2h^x_z\Big),
\label{eq_motion_v_03} 
\end{eqnarray}
\end{subequations}

\vspace*{-7mm}
\noindent
where the prime implies the derivative with respect to $r$. 
In the set of equations,  
the equations (\ref{eq_motion_v_01}) and 
(\ref{eq_motion_v_02}) imply (\ref{eq_motion_v_03}). 
On the other hand, 
in the Maxwell equation (\ref{maxwell_eq_01}), the 
$x$-component gives a nontrivial contribution,  
\begin{equation}
0=
A_x''+\frac{(r^3f)'}{r^3f}A_x'
+\frac{l^4}{r^4f^2}\Bigl(\omega^2-k^2f\Bigr)A_x
+\frac{2Q}{r^3f}{h^x_t}'. 
\label{eq_motion_v_04}
\end{equation}
Taking the limit in which the charge $q$ goes to zero, 
the metric and the gauge perturbations are 
completely decoupled. 
 
We now look for solutions of our set of equations.  
First of all, 
from the equations (\ref{eq_motion_v_01}) and (\ref{eq_motion_v_02}), 
we can obtain a second order differential equation for ${h^x_t}'(r)$ 
and $A_x(r)$,  
\begin{eqnarray}
0
&=&
{h^x_t}'''
+\frac{(r^9f)'}{r^9f}{h^x_t}''
+\frac{1}{r^4f}
\bigg(
5(r^3f)'+\frac{l^4}{f}\Big(\omega^2-k^2f\Big)
\bigg)
{h^x_t}'  
\nonumber 
\\
&&
+\frac{6q^2l^2}{Q}
\bigg(
\frac{A_x''}{r^5}+\frac{(r^{-1}f)'}{r^4f}A_x'
+\frac{l^4\omega^2}{r^9f^2}A_x
\bigg). 
\label{eq_motion_v_05}
\end{eqnarray}
Together with the equation of motion (\ref{eq_motion_v_04}), 
we treat ${h^x_t}'(r)$ and $A_x(r)$ as independent variables. 
Having the solutions for these, one can get one for ${h^x_z}'(r)$ 
by using the equation (\ref{eq_motion_v_02}). 
In order to solve these equations, 
we find it is useful to introduce linear combinations of the variables 
\begin{equation}
\Phi_\pm\equiv
-\frac{8b^4}{l^8}r^5{h^x_t}'
+\bigg(-\frac{3al^4}{4Qb^2}+\frac{C_\pm}{Q}r^2\bigg)A_x, 
\end{equation}
with constants 
$$
C_\pm
=(1+a)\pm\sqrt{(1+a)^2+3ab^2k^2}, 
$$
so that we can obtain second order ordinary differential equations  
in terms of these new variables.  
In fact, the equations of motion (\ref{eq_motion_v_04}) and 
(\ref{eq_motion_v_05}) could be rearranged as  
\begin{eqnarray}
0
&=&
{\Phi_\pm}''
+\frac{(r^{-1}f)'}{r^{-1}f}{\Phi_\pm}'
+\frac{l^4}{r^4f^2}\Big(\omega^2-k^2f\Big)\Phi_\pm
-\frac{l^8C_\pm}{4b^4r^6f}\Phi_\pm.
\label{master_eq_00}
\end{eqnarray}
In the chargeless limit, the two equations of motion (\ref{master_eq_00}) 
for $\Phi_+(r)$ and $\Phi_-(r)$ give decoupled ones  
for $A_x(r)$ and ${h^x_t}'(r)$, respectively.     

We will consider these equations of motion in
low frequency limit so-called hydrodynamic regime.  
In the hydrodynamic regime we could obtain the diffusion pole 
and the thermal conductivity  
from retarded Green functions. 

\subsection{Tensor type perturbation}

Next we shall focus on the tensor type perturbation. 
By considering the spin or by calculating directly, 
the metric perturbation is decoupled from the gauge perturbation. 
Thus independent variables are 
$$
h_{xy}(x)\ne 0, \qquad
h_{xx}(x)=-h_{yy}(x), \qquad
(\mbox{others})=0.
$$
A nontrivial equation of motion in (\ref{einstein_eq_bh_01}) is
coming from $(x,y)$ component. 
As we did in the vector type perturbation, 
it might be convenient to introduce new variable 
$h^x_y(r)=g^{(0)xx}h_{xy}(r)=(l^2/r^2)h_{xy}(r)$. 
We then get the following equation of motion:  
\begin{equation}
0=
{h^x_y}''
+\frac{(r^5f)'}{r^5f}{h^x_y}'
+\frac{l^4}{r^4f^2}
\Big(\omega^2-k^2f\Big)h^x_y. 
\label{eq_motion_hxy_01}
\end{equation}
An another equation of motion for $h_{xx}(r)=-h_{yy}(r)$ 
is as the same form
of (\ref{eq_motion_hxy_01}).
We use this equation of motion to study the shear viscosity 
in the hydrodynamic approximation.  

The equations (\ref{master_eq_00}) and (\ref{eq_motion_hxy_01}) 
can be rewritten as Schr{\"o}dinger-like equations through suitable 
field redefinitions. 
Their potentials were derived by Kodama and Ishibashi~\cite{ki}.

\section{Diffusion Pole in Hydrodynamic Regime}\label{sec:Diffusion}
\setcounter{equation}{0}
\setcounter{footnote}{0}

In the hydrodynamic regime, 
it is standard to introduce new dimensionless coordinate 
$u=r^2_+/r^2$ which is normalized by the outer horizon.   
In this coordinate system, the horizon and the boundary are 
located at $u=1$ and $u=0$, respectively. 
Defining the new variable 
$B(u)\equiv\displaystyle\frac{A_x(u)}{\mu}=\frac{l^4}{4Qb^2}A_x(u)$ 
where $\mu$ is the chemical potential given by (\ref{chemical_potential}), 
our basic equations (\ref{eq_motion_v_01})-(\ref{eq_motion_v_03}) 
and (\ref{eq_motion_v_04}) are rewritten in this new coordinate system: 
\begin{subequations}
\begin{eqnarray}
0
&=&
{h^x_t}''-\frac{1}{u}{h^x_t}'
-\frac{b^2}{uf}\Big(\omega k h^x_z+k^2h^x_t\Big)
-3auB', 
\label{eq_motion_v_001}
\\
0
&=&
kf{h^x_z}'+\omega{h^x_t}'-3a\omega uB, 
\label{eq_motion_v_002}
\\
0
&=&
{h^x_z}''+\frac{(u^{-1}f)'}{u^{-1}f}{h^x_z}'
+\frac{b^2}{uf^2}\Bigl(\omega^2h^x_z+\omega kh^x_t\Bigr),
\label{eq_motion_v_003} 
\\
0
&=&
B''+\frac{f'}{f}B'+\frac{b^2}{uf^2}\Bigl(\omega^2-k^2f\Bigr)B
-\frac{1}{f}{h^x_t}', 
\label{eq_motion_v_004}
\end{eqnarray}
\end{subequations}

\vspace*{-7mm}
\noindent
with 
$$
f(u)=(1-u)(1+u-au^2).   
$$
Here the prime now means the derivative with respect to $u$. 
The equation (\ref{master_eq_00}) may be also written 
down as 
\begin{subequations}
\begin{equation}
0=
{\Phi_\pm}''
+\frac{(u^2f)'}{u^2f}{\Phi_\pm}'
+\frac{b^2}{uf^2}\Big(\omega^2-k^2f\Big)\Phi_\pm
-\frac{C_\pm}{f}\Phi_\pm, 
\label{master_eq_01}
\end{equation}
for 
\begin{equation}
\Phi_\pm
=
\frac{1}{u}{h^x_t}'-3aB+\frac{C_\pm}{u}B. 
\end{equation}
\end{subequations}

\vspace*{-7mm}
\noindent
Getting the solution for $\Phi_\pm(u)$, one can access to solutions 
for ${h^x_t}'(u)$ and $B(u)$, 
\begin{subequations}
\begin{eqnarray} 
{h^x_t}'
&=&
u\Phi_-
+\frac{3a}{C_+-C_-}u^2\Bigl(\Phi_+-\Phi_-\Bigr)
-\frac{C_-}{C_+-C_-}u\Bigl(\Phi_+-\Phi_-\Bigr),
\label{h_xt} 
\\
B
&=&
\frac{1}{C_+-C_-}u\Bigl(\Phi_+-\Phi_-\Bigr).
\label{B}
\end{eqnarray}
\end{subequations}

\vspace*{-7mm}
\noindent
The constants $C_\pm$ could be expanded in this regime, 
\begin{equation}
\begin{array}{rcl}
C_+
&=&
2(1+a)+{\displaystyle \frac{3ab^2}{2(1+a)}}k^2 + {\cal O}(k^4), 
\\
C_-
&=&
-{\displaystyle \frac{3ab^2}{2(1+a)}}k^2 + {\cal O}(k^4). 
\end{array}
\end{equation}

First, let us consider the equation for $\Phi_-(u)$.  
Following the usual way to solve differential equations, 
we impose a solution 
as $\Phi_-(u)=(1-u)^\nu F_-(u)$ where $F_-(u)$ is a 
regular function at the horizon $u=1$. 
Substituting this form into the equation of motion,  
one can fix the parameter $\nu$ as 
$\nu=\pm i\omega/(4\pi T)$ 
where $T$ is the temperature defined by the equation (\ref{temp}). 
We here choose 
$$
\nu=-i\frac{\omega}{4\pi T}, 
$$
as the incoming wave condition. 

Now we are in the position to solve the equation of motion 
in the hydrodynamic regime. 
We start by introducing the following series expansion 
with respect to small $\omega$ and $k$: 
\begin{equation}
F_-(u)
=
F_0(u)+\omega F_1(u)+k^2G_1(u)+{\cal O}(\omega^2, \ \omega k^2), 
\end{equation}
where $F_0(u)$, $F_1(u)$ and $G_{1}(u)$ are determined by imposing 
suitable boundary conditions. 
In order to do the perturbative analysis, 
it might be convenient to rewrite the equation (\ref{master_eq_01}) for 
$\Phi_-(u)$ as, 
\begin{eqnarray}
0
&=&
\bigg(u^2(1-u)(1+u-au^2){F_-}'\bigg)'
\nonumber 
\\
&&
+i\omega\frac{2b}{2-a}u^2\big(1+u-au^2\big)F_-' 
+i\omega\frac{b}{2-a}u\big(2+3u-4au^2\big)F_-
\nonumber 
\\
&&
+\omega^2\frac{b^2}{(2-a)^2}
\frac{u}{1+u-au^2}
\nonumber 
\\
&&
\hspace*{10mm}
\times\bigg(
(2-a)^2+(1-a)(3-a)u
+(1-4a+a^2)u^2
-a(2-a)u^3
+a^2u^4
\bigg)F_-
\nonumber 
\\
&&
-k^2b^2u
\bigg(1-\frac{3a}{2(1+a)}u\bigg)F_-. 
\label{master_eq_phi-_01}
\end{eqnarray}
The solution can be then obtained recursively\footnote{
The derivation of the solutions is given in Appendix A. 
}.
The result is as follows: 
\begin{subequations}
\begin{eqnarray}
F_0(u)
&=&
C, \quad (\mbox{const.}), 
\\
F_1(u)
&=&
iCb\Bigg\{
\frac{1+2a-2a^2}
{
2\sqrt{\displaystyle 1+4a}
\big(2-a\big)
}
\Bigg(
\log
\left(
\frac{\displaystyle 1-\frac{1-2au}{\sqrt{1+4a}}}
{\displaystyle 1-\frac{1-2a}{\sqrt{1+4a}}}
\right)
-\log
\left(
\frac{\displaystyle 1+\frac{1-2au}{\sqrt{1+4a}}}
{\displaystyle 1+\frac{1-2a}{\sqrt{1+4a}}}
\right)
\Bigg)
\nonumber
\\
&&
\hspace*{10mm}
+1-\frac{1}{u}
+
\frac{
1
}{2\big(2-a\big)}\log
\bigg(
\frac{1+u-au^2}{2-a}
\bigg)
\Bigg\},
\\
G_1(u)
&=&
\frac{Cb^2}{2(1+a)}\bigg(-1+\frac{1}{u}\bigg). 
\end{eqnarray}
\end{subequations}

\vspace*{-7mm}
\noindent
All of the solutions should be regular at the horizon $u=1$ and 
the functions $F_1(u)$ and $G_1(u)$ should be vanished there. 
The constant of integration $C$ will be estimated later. 

Next, we shall study the equation for $\Phi_+(u)$. 
It might be useful to introduce new variable $\widetilde{\Phi}_+(u)$, 
\begin{equation}
\Phi_+
\equiv
\bigg(-\frac{3a}{2(1+a)}+\frac{1}{u}\bigg)\widetilde{\Phi}_+.
\end{equation}
In terms of new variable, the equation of motion (\ref{master_eq_01}) 
for $\Phi_+(u)$ becomes
\begin{equation}
0=
\widetilde{\Phi}_+''
+\frac{\displaystyle\bigg(\Big(1-\frac{3a}{2(1+a)}u\Big)^2f\bigg)'}
{\displaystyle\Big(1-\frac{3a}{2(1+a)}u\Big)^2f}\widetilde{\Phi}_+'
+\frac{b^2}{uf^2}
\bigg(\omega^2-k^2f\Big(1+\frac{3a}{2(1+a)}u\Big)\bigg)\widetilde{\Phi}_+.
\label{master_eq_+}
\end{equation}
Assuming again 
$\widetilde{\Phi}_+(u)=(1-u)^\nu \widetilde{F}(u)$ where 
$\widetilde{F}(u)$ is a regular function at $u=1$, the singularity 
might be extracted. 
The equation of motion (\ref{master_eq_+}) becomes 
\begin{eqnarray}
0
&=&
\bigg(\Big(1-u\Big)\Big(1+u-au^2\Big)
\Big(1-\frac{3a}{2(1+a)}u\Big)^2\widetilde{F}'\bigg)'
\nonumber 
\\
&&
+2i\omega\frac{b}{2-a}
\Big(1+u-au^2\Big)
\Big(1-\frac{3a}{2(1+a)}u\Big)^2\widetilde{F}'
\nonumber
\\
&&
+i\omega\frac{b}{2-a}
\bigg(\Big(1+u-au^2\Big)\Big(1-\frac{3a}{2(1+a)}u\Big)^2\bigg)'
\widetilde{F}
\nonumber 
\\
&&
+\frac{\omega^2b^2}{(2-a)^2}
\frac{\displaystyle\bigg(1-\frac{3a}{2(1+a)}u\bigg)^2}
{u(1+u-au^2)}
\nonumber 
\\
&&
\hspace*{10mm}
\times
\Bigg(
(2-a)^2+(1-a)(3-a)u
+(1-4a+a^2)u^2
-a(2-a)u^3
+au^4
\Bigg)\widetilde{F}
\nonumber 
\\
&&
-\frac{k^2b^2}{u}
\bigg(1+\frac{3a}{2(1+a)}u\bigg)
\bigg(1-\frac{3a}{2(1+a)}u\bigg)^2\widetilde{F},
\label{master_eq_phi+_01} 
\end{eqnarray}
where we used the incoming wave condition $\nu=-i\omega/(4\pi T)$ as 
same as before. 

We impose a perturbative solution as 
\begin{equation}
\widetilde{F}(u)
=\widetilde{F}_0(u)
+\omega\widetilde{F}_1(u)
+k^2\widetilde{G}_1(u)
+{\cal O}(\omega^2, \ \omega k^2), 
\end{equation}
and then we obtain the following result\footnote{
The detail is given in Appendix B.
}: 
\begin{subequations}
\begin{eqnarray}
\widetilde{F}_0(u)
&=&
\widetilde{C}, \quad (\mbox{const.}), 
\\
\widetilde{F}_1(u)
&\equiv&
\widetilde{C}\widetilde{H}(u) 
\nonumber 
\\
&=&
i\frac{\widetilde{C}b}{2-a}
\Bigg\{
\frac{27a^2}{1+4a}\Bigg(\frac{1-u}{2+2a-3au}\Bigg)
\nonumber 
\\
&&
\hspace*{15mm}
+\frac{1-10a-2a^2}{2(1+4a)^{3/2}}
\Bigg(
\log
\left(
\frac{\displaystyle 1-\frac{1-2au}{\sqrt{1+4a}}}
{\displaystyle 1-\frac{1-2a}{\sqrt{1+4a}}}
\right)
-\log
\left(
\frac{\displaystyle 1+\frac{1-2au}{\sqrt{1+4a}}}
{\displaystyle 1+\frac{1-2a}{\sqrt{1+4a}}}
\right)
\Bigg)
\nonumber 
\\
&&
\hspace*{15mm}
+\frac{1}{2}
\log\Bigg(
\frac{1+u-au^2}{2-a}
\Bigg)
\Bigg\}, 
\\
\widetilde{G}_1(u)
&\equiv&
\widetilde{C}\widetilde{J}(u) 
\nonumber 
\\
&=&
\widetilde{C}b^2
\Bigg\{
-\frac{9a^2(14+31a+8a^2)}
{2(1+a)(1+4a)(2-a)^2}
\Bigg(
\frac{1-u}{2+2a-3au}
\Bigg)
\nonumber 
\\
&&
\hspace*{11mm}
+
\frac{(1+a)
\Big(3a(2-a)(5+2a)-2(1+a)(1-10a-2a^2)\log(3a)\Big)}
{(2-a)^3(1+4a)^{3/2}}
\nonumber 
\\
&&
\hspace*{21mm}
\times
\Bigg(
\log
\left(
\frac{\displaystyle 1-\frac{1-2au}{\sqrt{1+4a}}}
{\displaystyle 1-\frac{1-2a}{\sqrt{1+4a}}}
\right)
-
\log
\left(
\frac{\displaystyle 1+\frac{1-2au}{\sqrt{1+4a}}}
{\displaystyle 1+\frac{1-2a}{\sqrt{1+4a}}}
\right)
\Bigg)
\nonumber 
\\
&&
\hspace*{11mm}
-\frac{4(1+a)^2}{(2-a)^3}
\log u \log\Big(1-u\Big)
\nonumber 
\\
&&
\hspace*{11mm}
-\frac{(1+a)
\Big(
9a(2-a)+2(1+a)(1+4a)\log(3a)
\Big)}{(2-a)^3(1+4a)}
\log
\left(
\frac{1+u-au^2}{2-a}
\right)
\nonumber 
\\
&&
\hspace*{11mm}
-\frac{54a^2(1+a)}{(2-a)^2(1+4a)}
\Bigg(
\frac{u\log u}{2+2a-3au}
\Bigg)
\nonumber 
\\
&&
\hspace*{11mm}
-\frac{4(1+a)^2}{(2-a)^3}
\Bigg(
\mbox{Li}_2(u)-\frac{\pi^2}{6}
\Bigg)
\nonumber 
\\
&&
\hspace*{11mm}
+\frac{2(1+a)^2}{(1+4a)^{3/2}(2-a)^3}
\nonumber 
\\
&&
\hspace*{14mm}
\times
\Bigg(
\Big(1-10a-2a^2+(1+4a)^{3/2}\Big)
\nonumber 
\\
&&
\hspace*{23mm}
\times
\Big(
\log u
\log
\left(
1-\frac{2au}{1-\sqrt{1+4a}}
\right)
+\log(3a)
\log
\left(
\frac{\displaystyle 1-\frac{2au}{1-\sqrt{1+4a}}}
{\displaystyle 1-\frac{2a}{1-\sqrt{1+4a}}}
\right)
\nonumber 
\\
&&
\hspace*{29mm}
+\mbox{Li}_2
\left(
\frac{2au}{1-\sqrt{1+4a}}
\right)
-\mbox{Li}_2
\left(
\frac{2a}{1-\sqrt{1+4a}}
\right)
\Big)
\nonumber 
\\
&&
\hspace*{17mm}
-\Big(
1-10a-2a^2-(1+4a)^{3/2}
\Big)
\nonumber 
\\
&&
\hspace*{23mm}
\times
\Big(
\log u
\log
\left(
1-\frac{2au}{1+\sqrt{1+4a}}
\right)
+\log(3a)
\log
\left(
\frac{\displaystyle 1-\frac{2au}{1+\sqrt{1+4a}}}
{\displaystyle 1-\frac{2a}{1+\sqrt{1+4a}}}
\right)
\nonumber 
\\
&&
\hspace*{29mm}
+\mbox{Li}_2
\left(
\frac{2au}{1+\sqrt{1+4a}}
\right)
-\mbox{Li}_2
\left(
\frac{2a}{1+\sqrt{1+4a}}
\right)
\Big)
\Bigg)
\Bigg\}, 
\end{eqnarray}
\end{subequations}

\vspace*{-7mm}
\noindent
where $\mbox{Li}_2(u)$ is the polylogalithm\footnote{
The polylogalithm appears from 
$$
\Big(\mbox{Li}_2(u)\Big)'=-\frac{\log(1-u)}{u}.
$$
Some values are given as, 
$\mbox{Li}_2(-1)=-\pi^2/12$, $\mbox{Li}_2(0)=0$ 
and $\mbox{Li}_2(1)=\pi^2/6$. 
}. 
It should be mentioned that the defined functions $\widetilde{H}(u)$ 
and $\widetilde{J}(u)$ in $\widetilde{F}_1(u)$ and $\widetilde{G}_1(u)$ 
are finite at the boundary $u=0$.     

Let us consider the integration constants $C$ and $\widetilde{C}$. 
These could be estimated in terms of boundary values of the fields 
$$
\lim_{u\rightarrow 0} h^x_t(u)=(h^x_t)^0, \quad 
\lim_{u\rightarrow 0} h^x_z(u)=(h^x_z)^0, \quad
\lim_{u\rightarrow 0} B(u)=(B)^0. 
$$
Taking a derivative of $\Phi_\pm(u)$ and 
using the equation of motion (\ref{eq_motion_v_001}), 
we can get relations   
$$
u^2\Phi_\pm'
-C_\pm uB'
=\frac{b^2}{f}\Bigl(\omega k h^x_z+k^2h^x_t\Bigr)
-C_\pm B. 
$$
We evaluate the equations above at the boundary, 
\begin{equation}
\lim_{u\rightarrow0}
\Bigl(u^2\Phi'_\pm-C_\pm uB'\Bigr)
=b^2\Bigl(\omega k(h^x_z{})^0+k^2(h^x_t)^0\Bigr)
-C_\pm (B)^0, 
\end{equation}
so that we may fix the constants $C$ and $\widetilde{C}$ from 
$\mp$ parts, respectively,   
\begin{subequations}
\begin{eqnarray}
C
&=&
\frac{\displaystyle b
\Bigl(\omega k(h^x_z)^0+k^2(h^x_t)^0\Bigr)+\frac{3ab}{2(1+a)}k^2(B)^0}
{\displaystyle i\omega-\frac{b}{2(1+a)}k^2}, 
\label{C}
\\
\widetilde{C}
&=&
\frac{\displaystyle -b^2\Bigl(\omega k (h^x_z)^0+k^2(h^x_t)^0\Bigr)
+\Bigl(2(1+a)+\frac{3ab^2}{2(1+a)}k^2\Bigl)(B)^0}
{1+\omega\widetilde{H}(0)+k^2\widetilde{J}(0)}, 
\end{eqnarray}
\end{subequations}

\vspace*{-7mm}
\noindent   
where we used the obtained solutions $\Phi_\pm(u)$ and the 
relation (\ref{B}) for $B'(u)$. 
It should be noted that the boundary value of $uB'(u)$ is 
vanished. 
In the equation (\ref{C}), one can see the existence of the 
hydrodynamic pole in the complex $\omega$-plane. 

Now we proceed to calculate the Minkowskian correlators.  
For the vector type perturbation, 
the on-shell action (\ref{on-shell_action_h_a}) becomes
\begin{eqnarray}
S[h^x_t, h^x_z, B]
=
\frac{l^3}{32\kappa^2b^4}
\!\int\!
\frac{\dd^4k}{(2\pi)^4}
\Biggl\{
&&
\frac{1}{u}h^x_t(-k, u){h^x_t}'(k, u)
-\frac{1}{u^2}h^x_t(-k, u)h^x_t(k, u)
\nonumber 
\\
&&
-\frac{f(u)}{u}h^x_z(-k, u){h^x_z}'(k, u)
+\frac{f(u)}{u^2}h^x_z(-k, u)h^x_z(k, u) 
\nonumber 
\\
&&
-3af(u)B(-k, u)\Bigl(B'(k, u)-\frac{1}{f(u)}h^x_t(k, u)\Bigr)
\Biggr\}
\Bigg|_{u=0}^{u=1}. 
\nonumber 
\\  
\end{eqnarray}
Using the obtained solutions, we can lead the following relations 
between the radial 
derivative of the fields and their boundary values near 
the boundary $u=\varepsilon$:  
\begin{subequations}
\begin{eqnarray}
{h^x_t}'(\varepsilon)
&=&
-b^2\Bigl(\omega k(h^x_z)^0+k^2(h^x_t)^0\Bigr)
\nonumber 
\\
&&
+\frac{\varepsilon}
{\displaystyle i\omega-\frac{b}{2(1+a)}k^2}
\Biggl\{
b\Bigl(\omega k(h^x_z)^0+k^2(h^x_t)^0\Bigr)
+3ia\omega(B)^0
+{\cal O}(\omega^2k, \omega k^2)
\Biggr\}
\nonumber 
\\
&&
+{\cal O}(\varepsilon^2), 
\\
{h^x_z}'(\varepsilon)
&=&
b^2\Bigl(\omega^2(h^x_z)^0+\omega k(h^x_t)^0\Bigr) 
\nonumber 
\\
&&
-\frac{\varepsilon}
{\displaystyle i\omega-\frac{b}{2(1+a)}k^2}
\Biggl\{
b\Bigl(\omega^2(h^x_z)^0+\omega k(h^x_t)^0\Bigr)
+\frac{3ab}{2(1+a)}\omega k(B)^0
+{\cal O}(\omega^2k, \omega k^2)
\Biggr\}
\nonumber 
\\
&&
+{\cal O}(\varepsilon^2), 
\\
B'(\varepsilon)
&=&
-
\frac{1}{\displaystyle i\omega-\frac{b}{2(1+a)}k^2}
\Biggl\{
\frac{b}{2(1+a)}\Bigl(\omega k(h^x_z)^0+k^2(h^x_t)^0\Bigr)
+i\frac{3a}{2(1+a)}\omega(B)^0 
\nonumber 
\\
&&
\hspace*{36mm}
-i\biggl(
i\omega-\frac{b}{2(1+a)}k^2
\biggr)
\frac{(2-a)^2b}{4(1+a)^2}\omega(B)^0
+{\cal O}(\omega^2k^2, k^4)
\Biggr\}
\nonumber 
\\
&&
+
\Bigl(
b^2k^2(B)^0
+{\cal O}(\omega k^2)
\Bigr)\log\varepsilon
+{\cal O}(\varepsilon). 
\end{eqnarray}
\end{subequations}

\vspace*{-7mm}
\noindent
By using the relation (\ref{green_function}) and 
the definition (\ref{diff_green_function}),  
we can read off the correlators in the hydrodynamic approximation, 
\begin{subequations}
\begin{eqnarray}
G_{xt \ xt}(\omega, k)
&=&
\frac{l^3}{16\kappa^2b^3}
\Biggl(
\frac{k^2}{i\omega-Dk^2}
\Biggr), 
\\
G_{xt \ xz}(\omega, k)
&=&
G_{xz \ xt}(\omega, k)
=
-\frac{l^3}{16\kappa^2b^3}
\Biggl(
\frac{\omega k}{i\omega-Dk^2}
\Biggr), 
\\
G_{xz \ xz}(\omega, k)
&=&
\frac{l^3}{16\kappa^2b^3}
\Biggl(
\frac{\omega^2}{i\omega-Dk^2}
\Biggr), 
\\
G_{xt \ x}(\omega, k)
&=&
G_{x \ xt}(\omega, k)
=
-\frac{2Q}{e^2l^3}
\Biggl(
\frac{i\omega}{i\omega-Dk^2}
\Biggr), 
\label{g_xtx}
\\
G_{xz \ x}(\omega, k)
&=&
G_{x \ xz}(\omega, k)
=
\frac{Qb}{(1+a)e^2l^3}
\Biggl(
\frac{\omega k}{i\omega-Dk^2}
\Biggr), 
\label{g_xzx}
\\
G_{x \ x}(\omega, k)
&=&
\frac{3al}{4(1+a)b^2e^2}
\Biggl(
\frac{i\omega}{i\omega-Dk^2}
\Biggr)
-\frac{(2-a)^2l}{8(1+a)^2be^2}i\omega, 
\label{g_xx} 
\end{eqnarray}
\end{subequations}

\begin{figure} [htbp]
\begin{minipage}{0.43\hsize}
\begin{center}
\includegraphics*[scale=0.55]{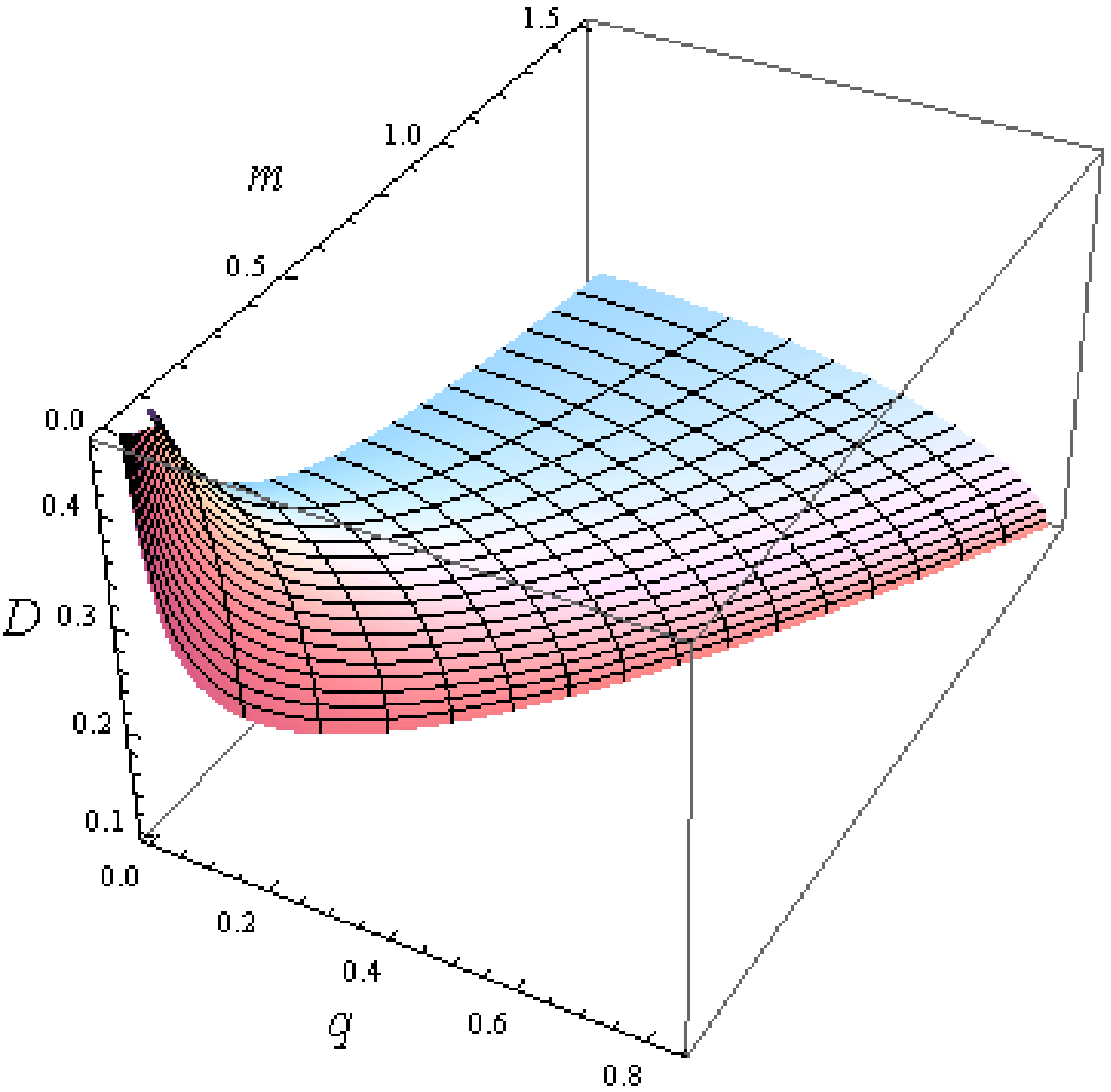}
\end{center}
\caption{$D$ vs. $q$ and $m$
 ($l=1$)}
\label{fig_d-q-m}
\end{minipage}
\qquad\quad
\begin{minipage}{0.43\hsize}
\vspace*{14mm}
\begin{center}
\includegraphics*[scale=0.55]{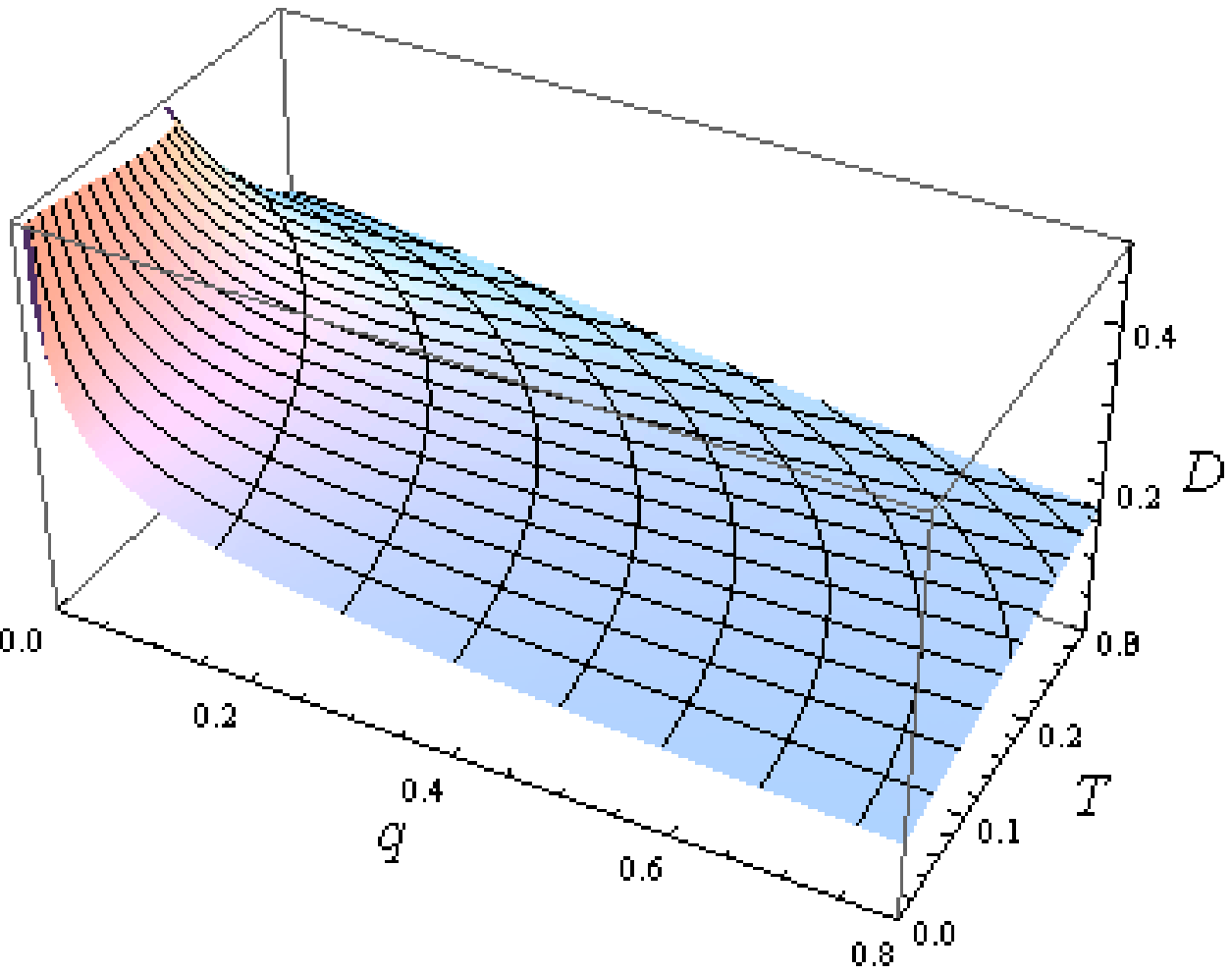}
\end{center}
\caption{$D$ vs. $q$ and $T$ $(l=1)$}
\label{fig_d-q-t}
\end{minipage}
\end{figure}

\vspace*{-7mm}
\noindent
where we subtracted the contact terms. 
In the final expression above 
we rescaled the gauge field $(B)^0$ to the original one 
 $(A_x)^0=\displaystyle\frac{4Qb^2}{l^4}(B)^0$ and raised 
and lowered the indices by using 
the flat Minkowski metric $\eta_{\mu\nu}=\mbox{diag}(-, +, +, +)$ in 
the four-dimensional boundary theory. 
Taking the limit in which the charge $q$ goes to zero, 
the results coincide with the known ones in \cite{pss}.  
In this limit, 
the correlators (\ref{g_xtx}) and (\ref{g_xzx}) vanish,  
while the correlator (\ref{g_xx}) has no diffusion pole and 
the subleading term reproduces the consistent result. 
The same interesting structure was found in 
the single $(1, 0, 0)$ $R$-charged black 
hole\cite{ss2}.     
The constant $D$ is the diffusion constant
\begin{equation}
D=\frac{b}{2(1+a)}
=\frac{1}{4}
\Biggl(
\frac{m^{5/3}}{3q^2}
\bigg(1+2\cos\Big(\frac{\theta}{3}+\frac{4}{3}\pi\Big)\bigg)
\Biggr)^{-\frac{3}{2}},
\label{diffusion_const}
\end{equation}
with
$$
\theta
=
\arctan
\Bigg(
\frac{3\sqrt{3}q^2\sqrt{\displaystyle 4m^3l^2-27q^4}}{2m^3l^2-27q^4}
\Bigg).
$$
All of the correlators in the vector type perturbation 
exhibit a diffusion pole. 
The behavior of the diffusion constant is drawn as a function of 
the charge $q$ and the mass $m$ in Figure \ref{fig_d-q-m} and 
as a function of the charge $q$ and the temperature $T$ in 
Figure \ref{fig_d-q-t}.   
In the chargeless limit, the diffusion constant becomes 
$$
D\rightarrow D_0=\frac{1}{4\pi T_0},  
$$
where the temperature $T_0$ is given in (\ref{temp_0}).

\section{Shear Viscosity in Hydrodynamic Regime}\label{sec:Viscosity}
\setcounter{equation}{0}
\setcounter{footnote}{0}

In this section, 
we solve the equation of motion (\ref{eq_motion_hxy_01}) 
in the hydrodynamic regime and obtain the shear viscosity. 
We could also see the hydrodynamic relation and 
the formulation of the thermal conductivity. 

After changing the coordinate $r$ to $u=r^2_+/r^2$,  
the equation (\ref{eq_motion_hxy_01}) can be rewritten as
\begin{equation}
0={h^x_y}''
+
\frac{(u^{-1}f)'}{u^{-1}f}
{h^x_y}'
+\frac{b^2}{uf^2}
\Big(
\omega^2-k^2f
\Big)
h^x_y, 
\label{eq_motion_xy}
\end{equation}
with 
$$
f(u)=(1-u)(1+u-au^2),  
$$ 
where the prime means the derivative with respect to $u$. 
Removing the singularity around $u=1$, the equation becomes
\begin{eqnarray}
0
&=&
\Big(
\frac{1}{u}
\big(1-u\big)
\big(1+u-au^2\big)
F'
\Big)'
\nonumber
\\
&&
+i\omega\frac{2b}{\big(2-a\big)}
\frac{1}{u}
\big(1-u\big)
\big(1+u-au^2\big)F'
-i\omega\frac{b}{\big(2-a\big)}
\frac{1}{u^2}\big(1+au^2\big)F
\nonumber
\\
&&
+\omega^2\frac{b^2}
{\big(2-a\big)^2
u^2\big(1+u-au^2\big)}
\nonumber
\\
&&
\hspace*{12mm}
\times
\Bigg(
\big(
a-2
\big)^2
+
\big(
a-3
\big)
\big(
a-1
\big)u
+
\big(
a^2
-4a
+1
\big)u^2
+a
\big(
a-2
\big)u^3
+
a^2u^4
\Bigg)
F
\nonumber
\\
&&
-k^2\frac{b^2}{u^2}F,
\label{eq_motion_hxy_00}
\end{eqnarray}
where we imposed the incoming wave condition
\begin{equation}
h^x_y(u)=(1-u)^{-i\omega/(4\pi T)}F(u).
\end{equation}
Perturbative solutions for $F(u)$,
\begin{equation}
F(u)=F_0(u)+\omega F_1(u)+k^2G_1(u)
+{\cal O}(\omega^2, \ \omega k^2),
\end{equation}
can be obtained as\footnote{
The detail is given in Appendix C.
}
\begin{subequations}
\begin{eqnarray}
F_0(u)
&=&
C, \quad (\mbox{const.}),
\\
F_1(u)
&\equiv&
CH(u)
\nonumber 
\\
&=&
i\frac{Cb}{2\big(2-a\big)}
\Bigg\{
-\frac{3}{\sqrt{1+4a}}
\Bigg(
\log
\left(
\frac{\displaystyle 1-\frac{1-2au}{\sqrt{1+4a}}}
{\displaystyle 1-\frac{1-2a}{\sqrt{1+4a}}}
\right)
-\log
\left(
\frac{\displaystyle 1+\frac{1-2au}{\sqrt{1+4a}}}
{\displaystyle 1+\frac{1-2a}{\sqrt{1+4a}}}
\right)
\Bigg)
\nonumber
\\
&&
\hspace*{21mm}
+\log\Bigg(
\frac{1+u-au^2}{2-a}
\Bigg)
\Bigg\},
\\
G_1(u)
&\equiv&
CJ(u) 
\nonumber 
\\
&=&
-\frac{Cb^2}
{\sqrt{1+4a}}
\Bigg\{\log
\left(
\frac{\displaystyle 1-\frac{1-2au}{\sqrt{1+4a}}}
{\displaystyle 1-\frac{1-2a}{\sqrt{1+4a}}}
\right)
-\log
\left(
\frac{\displaystyle 1+\frac{1-2au}{\sqrt{1+4a}}}
{\displaystyle 1+\frac{1-2a}{\sqrt{1+4a}}}
\right)
\Bigg\}.
\end{eqnarray}
\end{subequations}

\vspace*{-4mm}
\noindent
Since the function $h^x_y(u)$ goes to $(h^x_y)^0$ at the boundary $u=0$,
the constant $C$ can be fixed as
\begin{equation}
C=\frac{(h^x_y)^0}{1+\omega H(0)+k^2J(0)}. 
\end{equation}
Taking the limit $q\rightarrow 0$, the solution recovers
the result in \cite{pss}.
The solution of $h^x_x(u)$ is the same form as $h^x_y(u)$.

Let us evaluate the Minkowskian correlators. 
The relevant part of the metric perturbation in the on-shell action  
(\ref{on-shell_action_h_a}) becomes
\begin{eqnarray}
S[h^x_y, h^x_x, h^x_y]
=
-\frac{l^3}{32\kappa^2b^4}
\!\int\!\frac{\dd^4k}{(2\pi)^4}
\Bigg\{
&&
\frac{f(u)}{u}h^x_y(-k, u){h^x_y}'(k, u)
-\frac{f(u)}{u^2}h^x_y(-k, u){h^x_y}(k, u)
\nonumber
\\
&&
+
\frac{f(u)}{u}h^x_x(-k, u){h^x_x}'(k, u)
-\frac{f(u)}{u^2}h^x_x(-k, u)h^x_x(k, u)
\Bigg\}
\Bigg|_{u=0}^{u=1}.
\nonumber
\\
\label{on-shell_action_h1}
\end{eqnarray}
Near the boundary $u=\varepsilon$,
using the perturbative solution for $h^x_y(u)$,
we can obtain 
\begin{equation}
{h^x_y}'(\varepsilon)
=
\varepsilon b\Bigl(i\omega+bk^2\Bigr)(h^x_y)^0
-b^2k^2(h^x_y)^0
+{\cal O}(\omega^2, \omega k^2). 
\end{equation}
The same relation for $h^x_x(u)$ might be satisfied.
Therefore we can read off the correlation functions
from the on-shell action (\ref{on-shell_action_h1}),
\begin{eqnarray}
G_{xy \ xy}(\omega, k)
&=&
G_{xx \ xx}(\omega, k)
=
G_{yy \ yy}(\omega, k)
\nonumber
\\
&=&
-\frac{l^3}{16\kappa^2b^3}
\Big(i\omega+bk^2
\Big),
\end{eqnarray}
where we subtract contact terms.

\begin{figure} [htbp]
\begin{minipage}{0.43\hsize}
\begin{center}
\includegraphics*[scale=0.55]{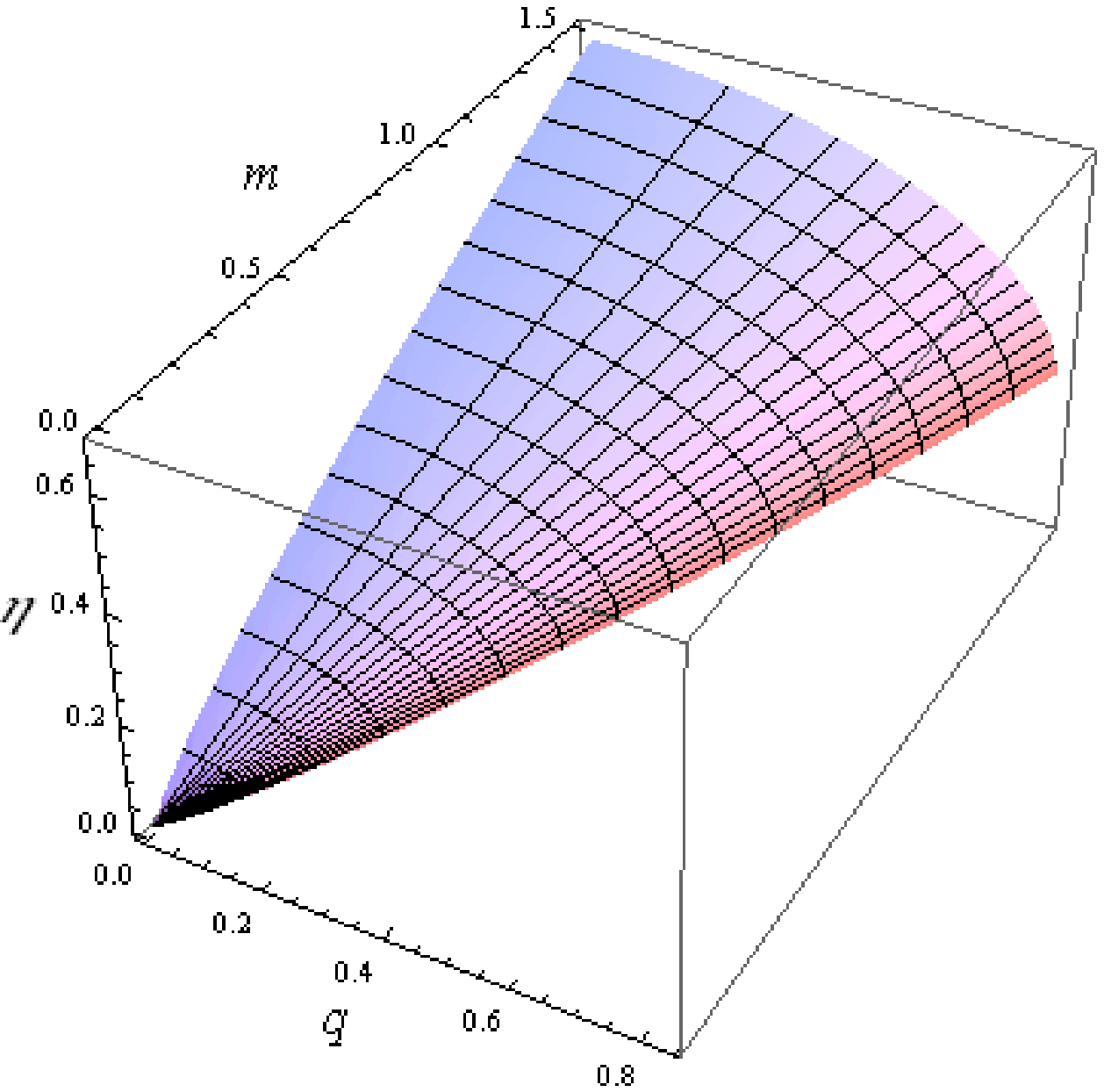}
\end{center}
\caption{$\eta$ vs. $q$ and $m$
 ($\kappa=l=1$)}
\label{fig_eta-q-m}
\end{minipage}
\qquad\quad
\begin{minipage}{0.43\hsize}
\vspace*{14mm}
\begin{center}
\includegraphics*[scale=0.55]{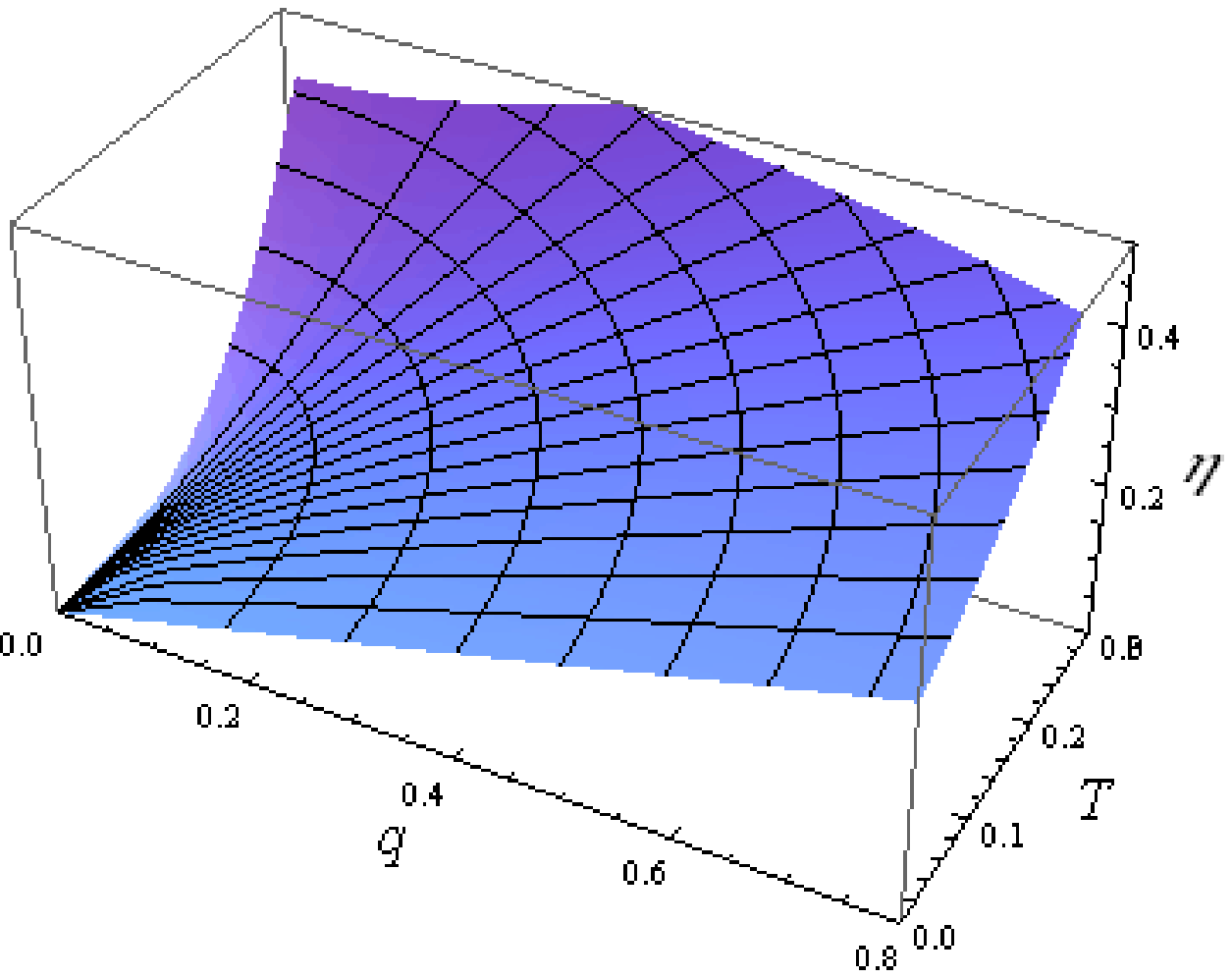}
\end{center}
\caption{$\eta$ vs. $q$ and $T$ $(\kappa=l=1)$}
\label{fig_eta-q-t}
\end{minipage}
\end{figure}

The result above can be used to estimate the shear viscosity $\eta$
via Kubo formula, 
\begin{equation}
\eta
=
-\lim_{\omega\rightarrow 0}
\frac{\mbox{Im} (G(\omega, 0))}{\omega}
=
\frac{l^3}{16\kappa^2b^3}. 
\label{viscosity}
\end{equation}
Therefore we can conclude the following relation
between the shear viscosity $\eta$ and the entropy density $s$ 
which is given in the equation (\ref{entropy}): 
\begin{equation}
\frac{\eta}{s}=\frac{1}{4\pi}.
\end{equation}
The behavior of the shear viscosity is drawn as a function of the 
charge $q$ and the mass $m$ in Figure \ref{fig_eta-q-m} and 
as a function of the charge $q$ and the temperature $T$ in Figure 
\ref{fig_eta-q-t}. 

\vspace*{2mm}

In hydrodynamics, the following relation is held: 
\begin{equation}
D=\frac{\eta}{\epsilon+p}, 
\end{equation}
where $\epsilon$ and $p$ are the energy density and the pressure defined 
in (\ref{energy}) and (\ref{pressure}), respectively.  
Using the obtained diffusion constant (\ref{diffusion_const}), 
the shear viscosity could be calculated. 
We can confirm the result coincides with (\ref{viscosity}) which 
was obtained from  Kubo formula.    

\begin{figure} [htbp]
\begin{minipage}{0.43\hsize}
\vspace*{15mm}
\begin{center}
\includegraphics*[scale=0.55]{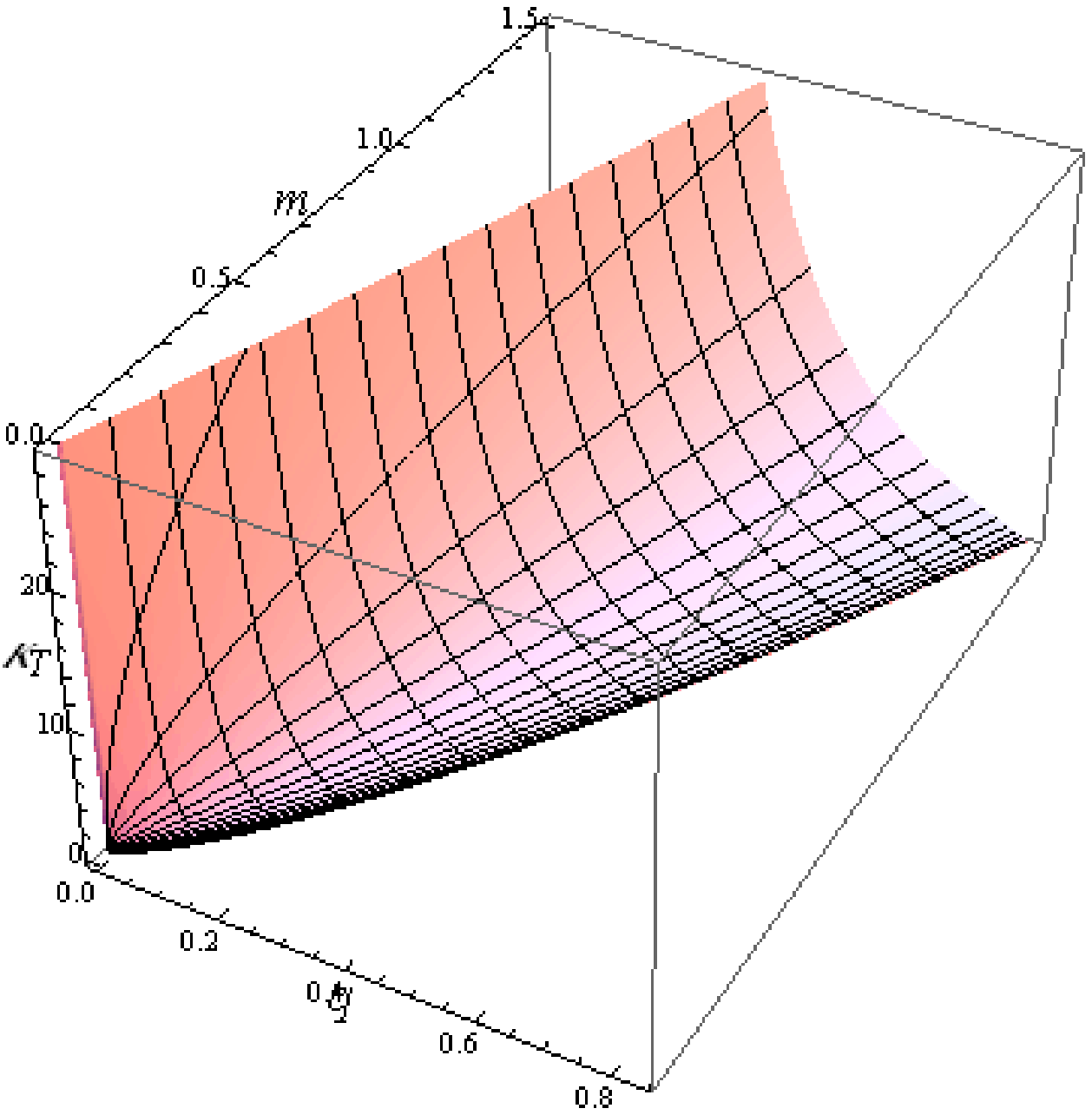}
\end{center}
\caption{$\kappa_T$ vs. $q$ and $m$
 ($\kappa=l=1$)}
\label{fig_kappa-q-m}
\end{minipage}
\qquad\quad
\begin{minipage}{0.43\hsize}
\vspace*{15mm}
\begin{center}
\includegraphics*[scale=0.55]{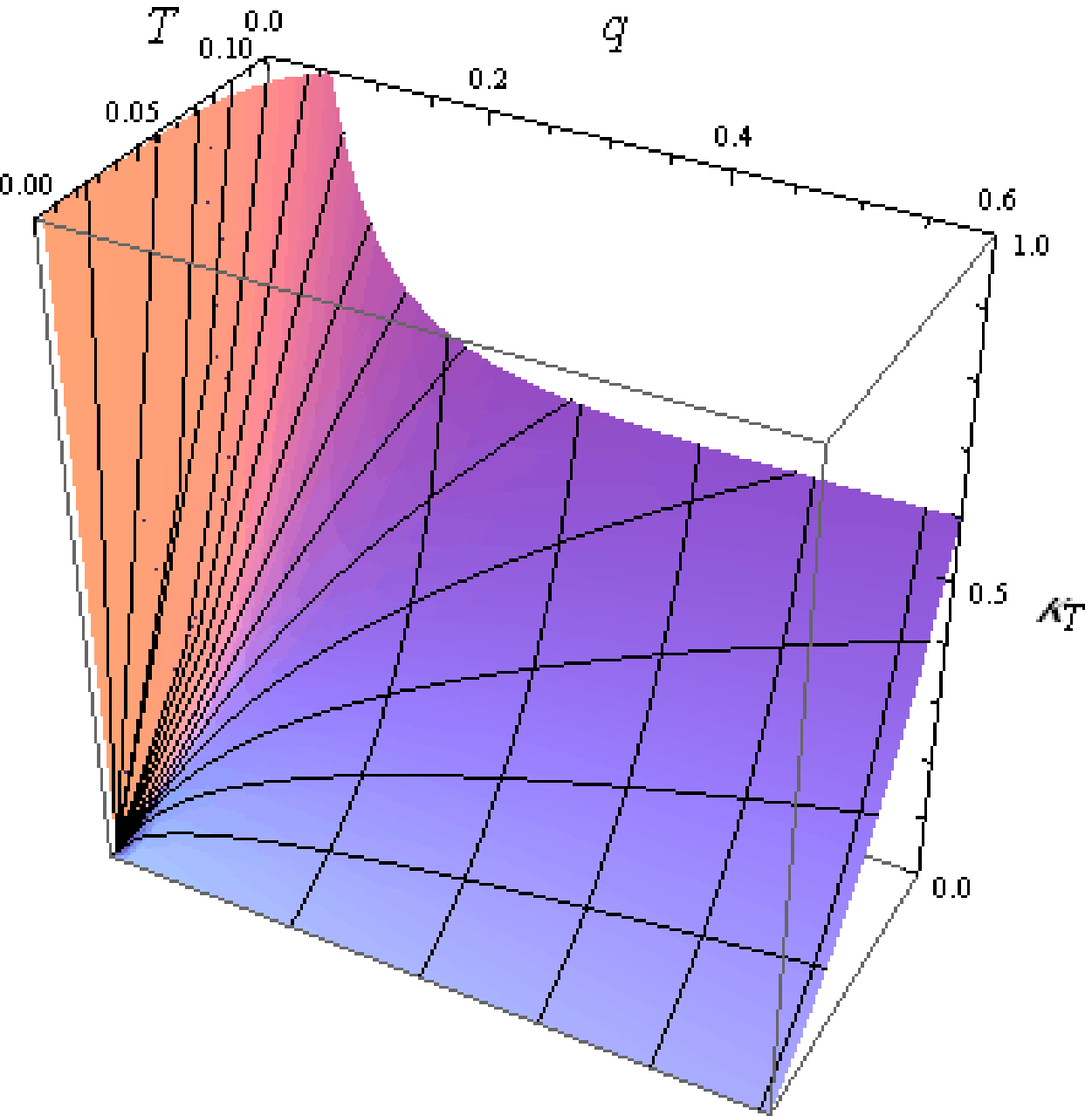}
\end{center}
\caption{$\kappa_T$ vs. $q$ and $T$ $(\kappa=l=1)$}
\label{fig_kappa-q-t}
\end{minipage}
\end{figure}

The thermal conductivity $\kappa_T$ can be also computed from 
the Green function by using Kubo formula~\cite{ss2}, 
\begin{equation}
\kappa_T
=
-\frac{(\epsilon+p)^2}{\rho^2T}\lim_{\omega\rightarrow 0}
\frac{\mbox{Im}(G(\omega, 0))}{\omega},    
\end{equation}
where the density of physical charge $\rho$ is given 
by (\ref{charge_density}).  
Here we can use the retarded Green function
$G_{x \ x}(\omega, 0)$ given by (\ref{g_xx}) as $G(\omega, 0)$.  
Thus we obtain
\begin{equation}
\kappa_T
=
2\pi^2
\Biggl(\frac{e^2l^2}{\kappa^2}\Biggr)
\frac{\eta T}{\mu^2}=2\pi^2
 \frac{N_c}{N_f} 
\frac{\eta T}{\mu^2}. 
\end{equation}
The behavior of the thermal conductivity $\kappa_T$ is drawn as a 
function of  
the charge $q$ and the mass in Figure \ref{fig_kappa-q-m} and 
as a function of the charge $q$ and the temperature $T$ in 
Figure \ref{fig_kappa-q-t}. 
%

\section{Conclusions and Discussions}\label{sec:Conclusion}
\setcounter{equation}{0}
\setcounter{footnote}{0}

In this paper we considered holographic QCD 
in the presence of the baryon density by introducing the bulk-filling 
branes.  
We use RN-AdS black hole geometry as the gravity dual of such system. 
We have seen the diffusion pole structure in vector type perturbation. 
It is worth mentioning that the correlator 
of Maxwell fields in the vector mode $G_{x \ x}(\omega, k)$ 
has the diffusion pole unlike the charge free case. 
The transport coefficients have been calculated 
in holographic hydrodynamics and their temperature and density 
dependence was demonstrated.  

The diffusion constant decreases as charge increases 
for fixed temperature. 
Physically, this implies that 
the fluid is less diffusible for 
large baryon density. 
By calculating the shear viscosity analytically, 
we showed that the shear viscosity $\eta$ and 
the entropy density $s$ satisfy the universal ratio 
$(\eta/s) =  1/(4\pi)$ which has been originally suggested in \cite{pss0}. 
For fixed temperature, the fluid becomes thicker as the charge increases.   
We have also seen that the diffusion constant and the shear viscosity 
satisfy the suitable relation for hydrodynamics.  
The calculation of the thermal conductivity 
shows that it satisfies (an analogue of) the Wiedemann-Franz low. 

It is very interesting to study 
the pole structure of scalar type as well as vector type of 
gravitational perturbations. 
Also it is important to carry out higher order calculations. 
Such result will be useful to get the higher order transport 
coefficients in the presence of the conserved current.  
We will report on these issues in the forthcoming publications.

In our interpretation, 
the fluctuations of bulk-filling branes are regarded
as master fields of the mesons. 
Near the horizon, the tension of the brane is zero due to 
the metric factor and it can lead to the long range fluctuation. 
This becomes the hydrodynamic mode.   
One important question is about the 
meaning of hydrodynamic mode in terms of meson physics. 

Further question in this direction is how we can understand 
the dispersion relations of vector modes of Maxwell fields 
in terms of the particle spectrum with dissipation.   
This vector mode cannot propagate in neutral medium while 
it can in charged medium. 
In addition, the tensor mode does not propagate in the medium.  
It is interesting to consider their interpretations in terms of 
meson physics. 
More thought on these points is to be pursued 
in the future. 
 
\vspace*{10mm}

\noindent
{\large{\bf Acknowledgments}}

\vspace*{2mm}
The work of SJS was supported by KOSEF Grant R01-2007-000-10214-0. 
This work is also supported by Korea Research Foundation Grant 
KRF-2007-314-C00052  and  SRC Program of the KOSEF through the 
CQUEST with grant number R11-2005-021. XHG, YM, FWS and TT would 
like to thank Hanyang University for warm hospitality.

\vspace*{10mm}

\noindent
{\large{\bf Appendix A. Perturbative solutions for $\Phi_-$}}

\renewcommand{\theequation}{A.\arabic{equation}}

\setcounter{equation}{0}
\setcounter{footnote}{0}

From the equation of motion (\ref{master_eq_phi-_01}), 
one can read off one for $F_0(u)$,  
\begin{equation}
0=\Big(
u^2
\big(
1-u
\big)
\big(
1+u-au^2
\big)
F_0'
\Big)'.
\end{equation}
A general solution is given by
\begin{eqnarray}
F_0(u)
&=&
C_0
+D_0
\Bigg\{
-\frac{1}{u}
+
\frac{1+2a-2a^2}
{2\sqrt{1+4a}
\big(
2-a
\big)}
\log
\left(
\frac{\displaystyle 1-\frac{1-2au}{\sqrt{1+4a}}}
{\displaystyle 1+\frac{1-2au}{\sqrt{1+4a}}}
\right)
\nonumber
\\
&&
\hspace*{19mm}
-\frac{1}{2-a}\log
\Big(
1-u
\Big)
+\frac{1}{2\big(2-a\big)}
\log
\Big(
1+u-au^2
\Big)
\Bigg\}.
\end{eqnarray}
Constants of integration $C_0$ and $D_0$ should be determined to be
a regular function at the horizon.
So we here choose $D_0=0$ and get
\begin{equation}
F_0(u)=C_0=C, \ ( \const).
\end{equation}

By using this solution, one can get an equation for $F_1(u)$ 
from (\ref{master_eq_phi-_01}), 
\begin{equation}
0=\Big(
u^2
\big(
1-u
\big)
\big(
1+u-au^2
\big)
F_1'
\Big)'
+i\frac{Cb}{2-a}
u\Big(
2+3u-4au^2
\Big).
\end{equation}
A general solution is
\begin{eqnarray}
F_1(u)
&=&
C_1
+
D_1
\Bigg(
\frac{1+2a-2a^2}
{2\sqrt{1+4a}\big(2-a\big)}
\log
\left(
\frac{\displaystyle 1-\frac{1-2au}{\sqrt{1+4a}}}
{\displaystyle 1+\frac{1-2au}{\sqrt{1+4a}}}
\right)
\nonumber
\\
&&
\hspace*{19mm}
-\frac{1}{u}
+\frac{1}{2\big(2-a\big)}
\log\Big(1+u-au^2\Big)
\Bigg)
\nonumber
\\
&&
+\frac{1}{2-a}
\Big(
iCb-D_1
\Big)
\log\Big(1-u\Big).
\end{eqnarray}
Again, removing the singularity at the horizon,
the constant $D_1$ should be
$$
D_1
=iCb.
$$
We also impose a boundary condition $F_1(u=1)=0$, so as to fix
the constant $C_1$,
$$
C_1
=
-iCb
\Bigg\{
\frac{1+2a-2a^2}
{2\sqrt{1+4a}\big(2-a\big)}
\log
\left(
\frac{\displaystyle 1-\frac{1-2a}{\sqrt{1+4a}}}
{\displaystyle 1+\frac{1-2a}{\sqrt{1+4a}}}
\right)
-1
+\frac{1}{2\big(2-a\big)}
\log\Big(2-a\Big)
\Bigg\}.
$$
Therefore the final form is
\begin{eqnarray}
F_1(u)
=
iCb
\Bigg\{
&&
\frac{1+2a-2a^2}
{2\sqrt{1+4a}\big(2-a\big)}
\Bigg(
\log
\left(
\frac{\displaystyle 1-\frac{1-2au}{\sqrt{1+4a}}}
{\displaystyle 1-\frac{1-2a}{\sqrt{1+4a}}}
\right)
-\log
\left(
\frac{\displaystyle 1+\frac{1-2au}{\sqrt{1+4a}}}
{\displaystyle 1+\frac{1-2a}{\sqrt{1+4a}}}
\right)
\Bigg)
\nonumber
\\
&&
+1-\frac{1}{u}
+\frac{1}{2\big(2-a\big)}
\log\Bigg(
\frac{1+u-au^2}{2-a}
\Bigg)
\Bigg\}.
\end{eqnarray}

A differential equation for $G_1(u)$ is
\begin{equation}
0=\Big(
u^2
\big(
1-u
\big)
\big(
1+u-au^2
\big)
G_1'
\Big)'
-
Cb^2u
\bigg(
1-\frac{3a}{2(1+a)}u
\bigg).
\end{equation}
A general solution is
\begin{eqnarray}
G_1(u)
&=&
\widetilde{C}_1
-\frac{\widetilde{D}_1}{u}
\nonumber
\\
&&
+\frac{
\Big(
1+2a-2a^2
\Big)
\Big(
Cb^2+2\widetilde{D}_1(1+a)
\Big)}{4\sqrt{1+4a}\big(1+a\big)\big(2-a\big)}
\log
\left(
\frac{\displaystyle 1-\frac{1-2au}{\sqrt{1+4a}}}
{\displaystyle 1+\frac{1-2au}{\sqrt{1+4a}}}
\right)
\nonumber
\\
&&
-\frac{Cb^2+2\widetilde{D}_1(1+a)}{2\big(1+a\big)\big(2-a\big)}
\bigg(
\log\Big(1-u\Big)
-\frac{1}{2}
\log\Big(
1+u-au^2
\Big)
\bigg),
\end{eqnarray}
and the constant $\widetilde{D}_1$ might be fixed as
$$
\widetilde{D}_1
=-\frac{Cb^2}{2(1+a)}.
$$
From the condition $G_1(u=1)=0$, we can fix the constant 
$\widetilde{C}_1$ as
$$
\widetilde{C}_1
=
-\frac{Cb^2}{2(1+a)}. 
$$
So we obtain the final form, 
\begin{equation}
G_1(u)
=
\frac{Cb^2}{2(1+a)}\bigg(-1+\frac{1}{u}\bigg). 
\end{equation}

\vspace*{5mm}

\noindent
{\large{\bf Appendix B. Perturbative solutions for $\Phi_+$}}

\renewcommand{\theequation}{B.\arabic{equation}}

\setcounter{equation}{0}
\setcounter{footnote}{0}

From the equation (\ref{master_eq_phi+_01}), 
we have a differential equation for $\widetilde{F}_0(u)$, 
\begin{equation}
0
=
\Bigg(\Big(1-u\Big)\Big(1+u-au^2\Big)
\Big(1-\frac{3a}{2(1+a)}u\Big)^2\widetilde{F}'_0
\Bigg)'. 
\end{equation}
A general solution is given by 
\begin{eqnarray}
\widetilde{F}_0(u)
&=&
C_0
-\frac{D_0}{2(2-a)^3}
\Bigg\{
\frac{18a(2-a)}{(1+4a)(2+2a-3au)}
-\frac{1-10a-2a^2}{(1+4a)^{3/2}}
\log
\left(
\frac{\displaystyle 1-\frac{1-2au}{\sqrt{1+4a}}}
{\displaystyle 1+\frac{1-2au}{\sqrt{1+4a}}}
\right)
\nonumber 
\\
&&
\hspace*{30mm}
+2\log\Big(1-u\Big)
-\log\Big(1+u-au^2\Big)
\Bigg\}. 
\end{eqnarray}
Since the function $\widetilde{F}_0$ should be regular at the horizon, 
we choose $D_0=0$ and get 
\begin{equation}
\widetilde{F}_0(u)
=C_0=\widetilde{C}, \quad (\mbox{const.}). 
\end{equation}

Substituting the solution to the equation (\ref{master_eq_phi+_01}), 
we get an equation for $\widetilde{F}_1(u)$, 
\begin{eqnarray}
0
&=&
\Bigg(
\Big(1-u\Big)\Big(1+u-au^2\Big)
\Big(1-\frac{3a}{2(1+a)}u\Big)^2\widetilde{F}_1'
\nonumber 
\\
&&
\hspace*{3mm}
+i\frac{\widetilde{C}b}{2-a}\Big(1+u-au^2\Big)
\Big(1-\frac{3a}{2(1+a)}u\Big)^2
\Bigg)'. 
\end{eqnarray}
A general solution is given as 
\begin{eqnarray}
\widetilde{F}_1(u)
&=&
C_1
+
\frac{2(1+a)^2}{(2-a)^3}D_1 
\Bigg\{
-\frac{18a(2-a)}{(1+4a)(2+2a-3au)}
\nonumber 
\\
&&
\hspace*{36mm}
+\frac{(1-10a-2a^2)}{(1+4a)^{3/2}}
\log
\left(
\frac{\displaystyle 1-\frac{1-2au}{\sqrt{1+4a}}}
{\displaystyle 1+\frac{1-2au}{\sqrt{1+4a}}}
\right)
\nonumber 
\\
&&
\hspace*{36mm}
+\log\Big(1+u-au^2\Big)
\Bigg\}
\nonumber 
\\
&&
+\frac{i\widetilde{C}(2-a)^2b-4D_1(1+a)^2}{(2-a)^3}
\log\Big(1-u\Big). 
\end{eqnarray}
The constant of integration $D_1$ should be 
$$
D_1
=
i\widetilde{C}\frac{(2-a)^2b}{4(1+a)^2}, 
$$
so that the singularity at the horizon would be removed. 
In addition, we require the condition $\widetilde{F}_1(u=1)=0$ 
to fix the constant $C_1$, 
$$
C_1
=
i
\frac{\widetilde{C}b}{2-a}
\Bigg\{
\frac{9a}{1+4a}
-\frac{1-10a-2a^2}{2(1+4a)^{3/2}}
\log
\left(
\frac{\displaystyle 1-\frac{1-2a}{\sqrt{1+4a}}}
{\displaystyle 1+\frac{1-2a}{\sqrt{1+4a}}}
\right)
-\frac{1}{2}\log\Big(2-a\Big)
\Bigg\}. 
$$
Then we get the final form of the solution 
\begin{eqnarray}
\widetilde{F}_1(u)
&=&
i\frac{\widetilde{C}b}{2-a}
\Bigg\{
\frac{27a^2}{1+4a}\Bigg(\frac{1-u}{2+2a-3au}\Bigg)
\nonumber 
\\
&&
\hspace*{15mm}
+\frac{1-10a-2a^2}{2(1+4a)^{3/2}}
\Bigg(
\log
\left(
\frac{\displaystyle 1-\frac{1-2au}{\sqrt{1+4a}}}
{\displaystyle 1-\frac{1-2a}{\sqrt{1+4a}}}
\right)
-\log
\left(
\frac{\displaystyle 1+\frac{1-2au}{\sqrt{1+4a}}}
{\displaystyle 1+\frac{1-2a}{\sqrt{1+4a}}}
\right)
\Bigg)
\nonumber 
\\
&&
\hspace*{15mm}
+\frac{1}{2}
\log\Bigg(
\frac{1+u-au^2}{2-a}
\Bigg)
\Bigg\}.
\end{eqnarray}

Similarly we have a differential equation for $\widetilde{G}_1(u)$, 
\begin{eqnarray}
0
&=&
\Bigg(
\Big(1-u\Big)\Big(1+u-au^2\Big)
\Big(1-\frac{3a}{2(1+a)}u\Big)^2\widetilde{G}_1'
\Bigg)'
\nonumber 
\\
&&
-\frac{\widetilde{C}b^2}{u}
\Big(1+\frac{3a}{2(1+a)}u\Big)
\Big(1-\frac{3a}{2(1+a)}u\Big)^2. 
\end{eqnarray}
A general solution of this equation can be obtained, 
\begin{eqnarray}
\widetilde{G}_1(u)
&=&
\widetilde{C}_1
\nonumber 
\\
&&
-\frac{9a\Big(\widetilde{D}_1+4\widetilde{C}(1+a)^2b^2\log(3a)\Big)}
{(2-a)^2(1+4a)}
\Bigg(\frac{1}{2+2a-3au}\Bigg)
\nonumber 
\\
&&
+\frac{\widetilde{C}(2-a)(1+4a)(14+15a+42a^2+14a^3)b^2
+6\widetilde{D}_1(1+a)(1-10a-2a^2)}
{12(2-a)^3(1+a)(1+4a)^{3/2}}
\nonumber 
\\
&&
\hspace*{10mm}
\times
\log
\left(
\frac{\displaystyle 1-\frac{1-2au}{\sqrt{1+4a}}}
{\displaystyle 1+\frac{1-2au}{\sqrt{1+4a}}}
\right)
\nonumber 
\\
&&
-\frac{\widetilde{C}b^2
\Big((2-a)(14+31a+8a^2)+24(1+a)^3\log(3a)\Big)
+6\widetilde{D}_1(1+a)}
{6(2-a)^3(1+a)}\log\Big(1-u\Big)
\nonumber 
\\
&&
-\frac{4\widetilde{C}(1+a)^2b^2}{(2-a)^3}
\log u\log \Bigl(1-u\Bigr)
\nonumber 
\\
&&
+\frac{\widetilde{C}(2-a)(14-21a-84a^2-76a^3)b^2
+6\widetilde{D}_1(1+a)(1+4a)}
{12(2-a)^3(1+a)(1+4a)}
\log\Big(1+u-au^2\Big)
\nonumber 
\\
&&
-\frac{54\widetilde{C}a^2(1+a)b^2}{(2-a)^2(1+4a)}
\Bigg(\frac{u\log u}{2+2a-3au}\Bigg)
\nonumber 
\\
&&
-\frac{4\widetilde{C}(1+a)^2b^2}{(2-a)^3}\mbox{Li}_2(u) 
\nonumber 
\\
&&
+\frac{2\widetilde{C}(1+a)^2b^2}{(2-a)^3(1+4a)^{3/2}}
\nonumber 
\\
&&
\hspace*{10mm}
\times
\Bigg\{
\Big(1-10a-2a^2+(1+4a)^{3/2}\Big)
\nonumber 
\\
&&
\hspace*{28mm}
\times
\bigg(
\log(3au)
\log
\left(
1-\frac{2au}{1-\sqrt{1+4a}}
\right)
+\mbox{Li}_2
\left(
\frac{2au}{1-\sqrt{1+4a}}
\right)
\bigg)
\nonumber 
\\
&&
\hspace*{17mm}
-\Big(
1-10a-2a^2-(1+4a)^{3/2}
\Big)
\nonumber 
\\
&&
\hspace*{28mm}
\times
\bigg(
\log(3au)
\log\left(
1-\frac{2au}{1+\sqrt{1+4a}}
\right)
+\mbox{Li}_2
\left(
\frac{2au}{1+\sqrt{1+4a}}
\right)
\bigg)
\Bigg\}. 
\nonumber 
\\
\end{eqnarray}
The constant of integration $\widetilde{D}_1$ might be fixed to 
remove the singularity $u=1$, 
$$
\widetilde{D}_1
=-
\frac{\widetilde{C}b^2}{6(1+a)}
\Big(
(2-a)(14+31a+8a^2)
+24(1+a)^3\log(3a)
\Big).
$$
Another constant of integration $\widetilde{C}_1$ is used to satisfy  
the condition $\widetilde{G}_1(u=1)=0$. 
The final expression of the solution is 
\begin{eqnarray}
\widetilde{G}_1(u)
=
\widetilde{C}b^2
\Bigg\{
&&
-\frac{9a^2(14+31a+8a^2)}
{2(1+a)(1+4a)(2-a)^2}
\Bigg(
\frac{1-u}{2+2a-3au}
\Bigg)
\nonumber 
\\
&&
+
\frac{(1+a)
\Big(3a(2-a)(5+2a)-2(1+a)(1-10a-2a^2)\log(3a)\Big)}
{(2-a)^3(1+4a)^{3/2}}
\nonumber 
\\
&&
\hspace*{30mm}
\times
\Bigg(
\log
\left(
\frac{\displaystyle 1-\frac{1-2au}{\sqrt{1+4a}}}
{\displaystyle 1-\frac{1-2a}{\sqrt{1+4a}}}
\right)
-
\log
\left(
\frac{\displaystyle 1+\frac{1-2au}{\sqrt{1+4a}}}
{\displaystyle 1+\frac{1-2a}{\sqrt{1+4a}}}
\right)
\Bigg)
\nonumber 
\\
&&
-\frac{4(1+a)^2}{(2-a)^3}
\log u\log\Big(1-u\Big)
\nonumber 
\\
&&
-\frac{(1+a)
\Big(
9a(2-a)+2(1+a)(1+4a)\log(3a)
\Big)}{(2-a)^3(1+4a)}
\log
\left(
\frac{1+u-au^2}{2-a}
\right)
\nonumber 
\\
&&
-\frac{54a^2(1+a)}{(2-a)^2(1+4a)}
\Bigg(
\frac{u\log u}{2+2a-3au}
\Bigg)
\nonumber 
\\
&&
-\frac{4(1+a)^2}{(2-a)^3}
\Bigg(
\mbox{Li}_2(u)-\frac{\pi^2}{6}
\Bigg)
\nonumber 
\\
&&
+\frac{2(1+a)^2}{(1+4a)^{3/2}(2-a)^3}
\nonumber 
\\
&&
\hspace*{3mm}
\times
\Bigg(
\Big(1-10a-2a^2+(1+4a)^{3/2}\Big)
\nonumber 
\\
&&
\hspace*{10mm}
\times
\Big(
\log u
\log
\left(
1-\frac{2au}{1-\sqrt{1+4a}}
\right)
+\log(3a)
\log
\left(
\frac{\displaystyle 1-\frac{2au}{1-\sqrt{1+4a}}}
{\displaystyle 1-\frac{2a}{1-\sqrt{1+4a}}}
\right)
\nonumber 
\\
&&
\hspace*{16mm}
+\mbox{Li}_2
\left(
\frac{2au}{1-\sqrt{1+4a}}
\right)
-\mbox{Li}_2
\left(
\frac{2a}{1-\sqrt{1+4a}}
\right)
\Big)
\nonumber 
\\
&&
\hspace*{8mm}
-\Big(
1-10a-2a^2-(1+4a)^{3/2}
\Big)
\nonumber 
\\
&&
\hspace*{13mm}
\times
\Big(
\log u
\log
\left(
1-\frac{2au}{1+\sqrt{1+4a}}
\right)
+\log(3a)
\log
\left(
\frac{\displaystyle 1-\frac{2au}{1+\sqrt{1+4a}}}
{\displaystyle 1-\frac{2a}{1+\sqrt{1+4a}}}
\right)
\nonumber 
\\
&&
\hspace*{18mm}
+\mbox{Li}_2
\left(
\frac{2au}{1+\sqrt{1+4a}}
\right)
-\mbox{Li}_2
\left(
\frac{2a}{1+\sqrt{1+4a}}
\right)
\Big)
\Bigg)
\Bigg\}. 
\end{eqnarray}

\vspace*{5mm}

\noindent
{\large{\bf Appendix C. Perturbative solutions for $h_{xy}$}}

\renewcommand{\theequation}{C.\arabic{equation}}

\setcounter{equation}{0}
\setcounter{footnote}{0}

From the equation of motion (\ref{eq_motion_hxy_00}), 
one can get an equation for $F_0(u)$, 
\begin{equation}
\Big(
\frac{1}{u}
\big(
1-u
\big)
\big(
1+u-au^2
\big)
F_0'
\Big)'
=0.
\end{equation}
A general solution is given by
\begin{eqnarray}
F_0(u)
=
C_0
+
\frac{D_0}{2-a}
\Bigg\{
&-&
\frac{3}
{2\sqrt{1+4a}}
\log
\left(
\frac{\displaystyle 1-\frac{1-2au}{\sqrt{1+4a}}}
{\displaystyle 1+\frac{1-2au}{\sqrt{1+4a}}}
\right)
\nonumber
\\
&-&
\log
\Big(
1-u
\Big)
+\frac{1}{2}
\log
\Big(
1+u-au^2
\Big)
\Bigg\}.
\end{eqnarray}
Constants of integration $C_0$ and $D_0$ should be determined to be
a regular function at the horizon.
So we here choose $D_0=0$ and get
\begin{equation}
F_0(u)=C_0=C, \ ( \const).
\end{equation}

By using this solution, one can get an equation for $F_1(u)$,
\begin{equation}
\Big(
\frac{1}{u}
\big(
1-u
\big)
\big(
1+u-au^2
\big)
F_1'
\Big)'
=
i\frac{Cb}{2-a}
\frac{1}{u^2}\Big(
1+au^2
\Big).
\end{equation}
A general solution is
\begin{eqnarray}
F_1(u)
&=&
C_1
-\frac{1}{2\big(2-a\big)^2}
\Big(iCb+\big(2-a\big)D_1
\Big)
\nonumber
\\
&&
\hspace*{15mm}
\times
\Bigg\{
\frac{3}{\sqrt{1+4a}}
\log
\left(
\frac{\displaystyle 1-\frac{1-2au}{\sqrt{1+4a}}}
{\displaystyle 1+\frac{1-2au}{\sqrt{1+4a}}}
\right)
-\log\Big(1+u-au^2\Big)
\Bigg\}
\nonumber
\\
&&
+\frac{1}{\big(2-a\big)^2}
\Big(
iCb\big(1-a\big)
-\big(2-a\big)D_1
\Big)
\log\Big(1-u\Big).
\end{eqnarray}
Removing the singularity at the horizon,
the constant $D_1$ should be
$$
D_1
=iC\frac{1-a}{2-a}b.
$$
We also impose a boundary condition $F_1(u=1)=0$, so as to fix
the constant $C_1$,
$$
C_1
=
i\frac{Cb}{2\big(2-a\big)}
\Bigg\{
\frac{3}{\sqrt{1+4a}}
\log
\left(
\frac{\displaystyle 1-\frac{1-2a}{\sqrt{1+4a}}}
{\displaystyle 1+\frac{1-2a}{\sqrt{1+4a}}}
\right)
-\log\Big(2-a\Big)
\Bigg\}.
$$
Therefore the final form is
\begin{eqnarray}
F_1(u)
=
i\frac{Cb}{2\big(2-a\big)}
\Bigg\{
&&
-\frac{3}{\sqrt{1+4a}}
\Bigg(
\log
\left(
\frac{\displaystyle 1-\frac{1-2au}{\sqrt{1+4a}}}
{\displaystyle 1-\frac{1-2a}{\sqrt{1+4a}}}
\right)
-\log
\left(
\frac{\displaystyle 1+\frac{1-2au}{\sqrt{1+4a}}}
{\displaystyle 1+\frac{1-2a}{\sqrt{1+4a}}}
\right)
\Bigg)
\nonumber
\\
&&
+\log\Bigg(
\frac{1+u-au^2}{2-a}
\Bigg)
\Bigg\}.
\end{eqnarray}

A differential equation for $G_1(u)$ is
\begin{equation}
\Big(
\frac{1}{u}
\big(
1-u
\big)
\big(
1+u-au^2
\big)
G_1'
\Big)'
=
\frac{Cb^2}{u^2}.
\end{equation}
The equation gives us the solution
\begin{eqnarray}
G_1(u)
&=&
\widetilde{C}_1
-\frac{C\big(1-2a\big)b^2
+3\widetilde{D}_1}
{2\sqrt{1+4a}\big(2-a\big)}
\log
\left(
\frac{\displaystyle 1-\frac{1-2au}{\sqrt{1+4a}}}
{\displaystyle 1+\frac{1-2au}{\sqrt{1+4a}}}
\right)
\nonumber
\\
&&
+\frac{Cb^2-\widetilde{D}_1}
{2\big(2-a\big)}
\bigg(
2\log\Big(1-u\Big)
-\log\Big(1+u-au^2\Big)
\bigg),
\end{eqnarray}
and the constant $\widetilde{D}_1$ might be fixed as
$$
\widetilde{D}_1
=Cb^2.
$$
We can also fix the constant $\widetilde{C}_1$ as
$$
\widetilde{C}_1
=
\frac{Cb^2}{\sqrt{1+4a}}
\log
\left(
\frac{\displaystyle 1-\frac{1-2a}{\sqrt{1+4a}}}
{\displaystyle 1+\frac{1-2a}{\sqrt{1+4a}}}
\right).
$$
Then we obtain the result,
\begin{eqnarray}
G_1(u)
&=&
-\frac{Cb^2}
{\sqrt{1+4a}}
\Bigg\{\log
\left(
\frac{\displaystyle 1-\frac{1-2au}{\sqrt{1+4a}}}
{\displaystyle 1-\frac{1-2a}{\sqrt{1+4a}}}
\right)
-\log
\left(
\frac{\displaystyle 1+\frac{1-2au}{\sqrt{1+4a}}}
{\displaystyle 1+\frac{1-2a}{\sqrt{1+4a}}}
\right)
\Bigg\}.
\end{eqnarray}

\vspace*{5mm}


\end{document}